\documentclass[rmp,aps,nofootinbib,twocolumn]{revtex4}

\usepackage{graphicx}
\usepackage{bm}

\newcommand{\bgamma}{{\bm{\gamma}}}
\newcommand{\bsigma}{{\bm{\sigma}}}

\begin{document}

\title
{Patterns and Collective Behavior in Granular Media: Theoretical
Concepts}

\author{Igor  S.~Aranson }
\email{aronson@msd.anl.gov} \affiliation{Materials Science
Division, Argonne National Laboratory, 9700 S. Cass Av, Argonne,
IL 60439}
\author{Lev S.~Tsimring} \email{ltsimring@ucsd.edu}
\affiliation{Institute for Nonlinear Science, University of
California, San Diego, 9500 Gilman Drive, La Jolla, CA 92093 }

\begin{abstract}
Granular materials are ubiquitous in our daily lives. While they
have been a subject of intensive engineering research for
centuries, in the last decade granular matter attracted
significant attention of physicists. Yet despite a major efforts
by many groups, the theoretical description of granular systems
remains largely a plethora of different, often contradicting
concepts and approaches. Authors give an overview of various
theoretical models emerged in the physics of granular matter, with
the focus on the onset of collective behavior and pattern
formation. Their aim is two-fold: to identify general principles
common for granular systems and other complex non-equilibrium
systems, and to elucidate important distinctions between
collective behavior in granular and continuum pattern-forming
systems.
\end{abstract}

\date{\today}
\maketitle
\tableofcontents

 \section{Introduction}
\subsection{Preliminary remarks}
\label{sec:intro}

Granular materials are ubiquitous in our daily lives and basic to
many industries.  Yet understanding their dynamic behavior remains
a major challenge in physics, see for review
\textcite{Duran:1999,Jaeger:1996,Kadanoff:1999,deGennes:1999,Gollub:1999,
Nedderman:1992,Ristow:1999,Rachenbach:2000,Ottino:2000}.
Granular materials are collections of discrete macroscopic solid
grains with sizes large enough that Brownian motion is irrelevant
(energy of 1 mm grain moving with typical velocity of 1 cm/sec
exceeds the thermal energy at least by 10 orders of magnitude).
Since thermodynamic fluctuations do not play a role, for granular
systems to remain active they have to gain energy either from shear or
vibration and are thus far from equilibrium. External volume
forces (gravity, electric and magnetic fields) and flows of
interstitial fluids such as water or air may also be used to
activate the grains. When subjected to a large enough driving
force, a granular system may exhibit a transition from a granular
solid to a liquid and various ordered patterns of grains may
develop. Understanding fundamentals of granular materials draws
upon and gives insights into many fields at the frontier of modern
physics: plasticity of solids, fracture and friction; complex
systems from equilibrium such as colloids, foams, suspensions, and
biological self-assembled systems. Moreover, particulate flows are
central to a large number of industries including the chemical,
pharmaceutical, food, metallurgical, agricultural and construction
industries. Beyond these industrial applications, particle
laden-flows are widespread in nature, for example dune migration,
erosion/deposition processes, landslides, underwater gravity
currents and coastal geomorphology, etc.

From a theoretical point of view, it is sometimes useful to employ
an analogy between granular matter and ordinary condensed matter
and to regard the grains as the equivalent of (classical) atoms.
However, this analogy is far from complete, since the dissipative
nature of grain interactions is the source of many differences
between the two ``kinds'' of matter. In particular, dissipation is
responsible for the fact that most states of granular matter are
metastable. The typical macroscopic size of the grains renders
thermal fluctuations negligible and most standard thermodynamic
concepts inapplicable. Whereas the behavior of dilute granular
systems (rapid granular gases) can often be explained using the
framework of kinetic theory (see e.g.
\textcite{Brilliantov:2004}), the quantitative theory of dense
granular assemblies is far less developed.

In recent years several comprehensive reviews and monographs
have appeared on the subject of granular physics, see
\cite{Jaeger:1996,Duran:1999,Ristow:2001,Aradian:2002,Rachenbach:2000,Brilliantov:2004,Kudrolli:2004,Ottino:2000}.
Yet in most of them the focus has been on actual phenomena and
experiments rather than on theoretical concepts and approaches to
the problems of granular physics. Furthermore, the scope of
granular physics has become so broad that we chose to limit
ourselves with reviewing the recent progress in a subfield of
granular {\em pattern formation} leaving out many interesting and
actively developing subjects. We loosely define pattern
formation as a dynamical process leading to the spontaneous
emergence of nontrivial spatially non-uniform structure which is
weakly dependent on initial and boundary conditions. According to
our working definition, we include in the scope of the review the
patterns in thin layers of vibrated grains (Sec. IV,V), patterns
in gravity-driven flows (Sec. VI), granular stratification and
banding (Sec. VII), as well as a multitude of patterns found in
granular assemblies with complex interactions (Sec. VIII). Before
delving into details of theoretical modelling of these
pattern-forming systems, we present a brief overview of the
relevant experimental findings and main theoretical concepts (Sec.
II and III).

\subsection{Fundamental microscopic interactions}
\label{subsec:fun_inter}

Probably the most fundamental microscopic property of granular
materials is irreversible energy dissipation  in the course of
interaction (collision) between the particles. For the case of
so-called dry granular materials, i.e.  when the interaction with
interstitial fluid such as air or water is negligible, the
encounter between grains results in dissipation of energy while
total mechanical momentum is conserved. In contrast to the
interaction of particles in molecular gases, the collisions of
macroscopic grains is generally {\it inelastic}. There are several
well-accepted models addressing the specifics of energy
dissipation in the course of collision, see for details e.g.
\textcite{Brilliantov:2004}. The simplest case corresponds to
in-deformable (hard) frictionless particles with fixed restitution
coefficient $0<e<1$ characterizing the fraction of energy lost in
the course of collision. The relation between the velocities after
the collisions (${\bf v}_{1,2}^\prime$) and before the collision
(${\bf v}_{1,2}$) for two identical spherical particles is given
by
\begin{equation}
{\bf v}_{1,2}^\prime={\bf v}_{1,2} \mp \frac{1+e}{2}[{\bf
n}_{12}({\bf v}_{1}-{\bf v}_{2})]{\bf n}_{12}. \label{collis}
\end{equation}
Here $ {\bf n}_{12}$ is the unit vector pointed from the center of
particle 1 to the center of particle 2 at the moment of collision.
The case of $e=1$ corresponds to the elastic collisions (particles
exchange their velocities) and $e=0$ characterizes fully inelastic
collisions. For $0<e<1$ the total energy loss is of the form
$$
\Delta E=-\frac{1-e^2}{4}|{\bf n}_{12}({\bf v}_{1}-{\bf v}_{2})|^2.
$$
Modelling collisions between particles by a fixed
restitution coefficient is very simple and intuitive, however this
approximation can be questionable in certain cases. For example,
approximation of granular media by a gas of hard particles with
fixed $e$ often yields non-physical behavior such as inelastic
collapse \cite{McNamara:1996}: divergence of the number of
collisions in a finite time, see Subsec. \ref{sec:cooling}. In
fact, the restitution coefficient is known to depend on the
relative velocities of colliding particles and approaches unity as
$|{\bf v}_1-{\bf v}_2|\to 0$. This dependence is captured by the
visco-elastic modelling of particle collision (see e.g.
\textcite{Ramirez:1999}). For non-spherical grains the restitution
coefficient may also depend on the point of contact
\cite{Goldsmith:1964}.

Tangential friction forces play an important role in the dynamics
of granular matter, especially in dense systems. Friction forces
are hysteretic and history dependent (the contact between two
grains can be either stuck due to dry friction or sliding
depending on the history of interaction).  This strongly nonlinear
behavior makes the analysis of frictional granular materials
extremely difficult. In the majority of theoretical studies, the
simplest Coulomb law is adopted: friction is independent on
sliding velocity as long as tangential force exceeds the certain
threshold \cite{Walton:1993}.  However, the main problem is
represented by calculation of the static friction forces.  It is well
known that frictional contact forces among solid particles exhibit
indeterminacy in case of multiple contacts per particle  because
there are less force balance constraints than stress components
(see, e.g. \textcite{McNamara:2004,Unger:2005}).  To resolve this
indeterminacy in simulations, various approximate algorithms have
been proposed.  In soft particle molecular dynamics simulations
the most widely used approach to calculating friction forces is
the spring-dashpot model \cite{Cundall:1979,Schaffer:1996}.
Another approach is taken in the contact dynamics method. By
assuming that all particles are rigid and treating all contacting
particles as performing instantaneous collisions (even those which
are in fact in persistent contact), one can compute the contact
forces generated during these collisions based on local force
balance and impenetrability of the particles constraints (see
\textcite{Moreau:1994,Brendel:2004}).

Viscous drag forces due to interaction with interstitial fluid
often affect the dynamics of granular materials.
Gas-driven particulate flows is an active research area in
the engineering community, see e.g. \textcite{Jackson:2000}.
Fluid-particle interactions are also involved in many geophysical
processes, e.g. dune formation \cite{Bagnold:1954}. Whereas
interaction of small individual particles with the fluid is
well-understood in terms of Stokes law, collective interaction and
mechanical momentum transfer from particles to fluid remains an open
problem. Various phenomenological constitutive equations are
used in the engineering community to model fluid-particulate flows,
see e.g. \textcite{Duru:2002}.

Finally, small particles can acquire electric charge of magnetic
moment. In this situation fascinating collective behavior emerge
due to competition between short-range collisions and long-range
electromagnetic forces, see e.g.
\textcite{Aranson:2000ed,Blair:2003,Sapozhnikov:2003}. Effects of
complex inter-particle interactions on pattern formation in granular
systems will be discussed in Sec.
\ref{sec:comp}.

\section{Overview of dynamic behavior in
granular matter} \label{sec:general}

In this Section we give a brief overview of the main experiments illustrating the
dynamical behavior of granular media and the phenomena to be discussed
in greater depth in the following Sections.  We classify the
experiments according to the way energy is injected  into the
system: vibration, gravity, or shear.

\subsection{Pattern formation in vibrated layers}
\label{subsec:vibr}

Quasi-two-dimensional sub-monolayers of grains subjected to
vertical vibration exhibit a surprizing bimodal regime
characterized by a dense cluster of closely packed  almost
immobile grains surrounded by gas of agitated particles,
\cite{Olafsen:1998}, Fig. \ref{fig_olafsen}. This clustering
transition occurs when the magnitude of vibration is reduced (the
system is ``cooled down'') which is reminiscent to the clustering
instability observed in non-driven (freely cooling) gas of
inelastic particles discovered by \textcite{Goldhirsch:1993}, Fig.
\ref{fig_goldhirsch}. Detailed consideration of clustering
phenomena in sub-monolayer systems is given in Sec.
\ref{sec:submon}.

\begin{figure}[ptb]
\caption{Top view  of dense immobile cluster coexisting with
dilute granular gas,  from \textcite{Olafsen:1998}.}
\label{fig_olafsen}
\end{figure}

\begin{figure}[ptb]
\caption{A typical clustering configuration in two dimensions,
restitution coefficient 0.6, number of particles 40,000, from
\textcite{Goldhirsch:1993}.} \label{fig_goldhirsch}
\end{figure}

Multilayers of granular materials subject to vertical vibration
exhibit spectacular pattern formation. In a typical experimental
realization a layer of granular material about 10-30 particle
diameters thick is energized by precise vertical vibration
produced by an electromagnetic shaker.  Depending on experimental
conditions, plethora of patterns can be observed, from stripes and
squares to hexagons and interfaces, see Fig. \ref{fig_vibr}. While
the first observations of patterns in vibrated layers were made
more than two centuries ago by \textcite{Chladni:1787} and
\textcite{Faraday:1831}, the current interest in these problems
was initiated by \textcite{Douady:1989,Fauve:1989} and culminated
in the discovery by \textcite{Umbanhowar:1996} of a remarkable
localized object, oscillon, Fig. \ref{fig_swinney_oscillon}.
Detailed consideration of these observations and their modelling
efforts is given in Sec. \ref{sec:multil}.

\begin{figure}[ptb]
\caption{Representative patterns in vertically vibrated granular
layers for various values of frequency and amplitude of the
vibration: stripes, squares, hexagons, spiral, interfaces,
 and localized oscillons,
from \textcite{Umbanhowar:1996}.} \label{fig_vibr}
\end{figure}

\begin{figure}[ptb]
\caption{Localized oscillon in vertically vibrate granular layer,
from \cite{Umbanhowar:1996}.} \label{fig_swinney_oscillon}
\end{figure}

In another set of experiments pattern formation was studied in a
horizontally vibrated system, see e.g.
\textcite{Ristow:1997,Tennakoon:1998,Liffman:1997}. While there
are certain common features, such as sub-harmonic regimes and
instabilities, horizontally vibrated systems do not show richness
of behavior typical for the vertically vibrated systems, and
nontrivial flow regimes are typically localized near the walls.
When the granular matter is polydisperse, vertical or horizontal shaking
often leads to segregation. The most well-known manifestation of this
segregation is the so-called ``Brazil nut" effect when large particles float
to the surface of a granular layer under vertical shaking \cite{rosato87}.
Horizontal shaking is also known to produce interesting segregation band patterns
oriented orthogonally to the direction of shaking
\cite{Mullin:2000,Mullin:2002} (see Fig. \ref{fig_mullin}).

\begin{figure}[ptb]
\caption{Sequence of snapshots of a layer of copper balls/poppy
seeds mixture in a horizontally shaken cavity (frequency 12.5 Hz,
amplitude 2 mm) at times 5 min, 10 min, 15 min, 30 min,  1 h, 6 h,
from \textcite{Mullin:2000}.} \label{fig_mullin}
\end{figure}

\subsection{Gravity-driven granular flows}
\label{subsection:pgdf}

Gravity-driven systems such as chute flows and  sandpiles often
exhibit nontrivial patterns and spatio-temporal structures.
Possibly the most spectacular are avalanches observed in the
layers of granular matter if the inclination exceeds the critical
angle (static angle of repose). Avalanches were a subject of
continued research for many decades, however only recently it was
established that the avalanche shape depends sensitively on the
thickness of the layer and the inclination angle: triangular
downhill avalanches in thin layers and balloon-shaped avalanches
in thicker layers which expand both uphill and downhill, see Fig.
\ref{fig_daerr1} and \textcite{Daerr:1999,Daerr:2001}.
Gravity-driven granular flows are prone to a variety of
non-trivial secondary instabilities in granular chute flow:
fingering \cite{Pouliquen:1997}, see Fig. \ref{fig_pouliquen97b},
longitudinal vortices in rapid chute flows
\cite{Forterre:2001,Borzsonyi:2005}, see Fig. \ref{fig_forterre},
long modulation waves \cite{Forterre:2003}, and others.

\begin{figure}
\caption{Sequence of  images illustrating evolution of avalanches
in thin layers on incline. Three images on left: triangular
avalanche in thin layer, point ${\bf b}$ in Fig. \ref{fig_daerr2}.
Three right images: up-hill avalanche in thicker layer, point
${\bf c}$ in Fig. \ref{fig_daerr2}, from \textcite{Daerr:1999}}
\label{fig_daerr1}
\end{figure}

\begin{figure}[ptb]
\caption{Fingering instability in chute flow. (a) Schematics of
the instability mechanism, the arrows represent trajectory of
coarse particles. Images  taken from front (b) and bottom (c)
illustrating accumulation of coarse particles between the
advancing fingers, from \textcite{Pouliquen:1997}.}
\label{fig_pouliquen97b}
\end{figure}

\begin{figure}[ptb]
\caption{Development of longitudinal vortices in the rapid
granular flow down rough incline, from \textcite{Forterre:2001}.}
\label{fig_forterre}
\end{figure}


Rich variety of patterns and instabilities has also been found in
underwater flows of granular matter: transverse instability of an
avalanche fronts, fingering, pattern formation in the sediment
behind the avalanche, etc. (see
\textcite{Daerr:2003,Malloggi:2005,Malloggi:2005b}). Whereas
certain pattern forming mechanisms are specific to the
water-granulate interaction, one also finds striking similarities
with the behavior of ``dry'' granular matter.

\subsection{Flows in rotating cylinders}
\label{subsection:cyl}

Energy is often supplied into a granular system through the shear
which is  driven by the moving walls of the container.  One of the
most commonly used geometries for this class of systems is a
horizontal cylinder rotated around its axis, or rotating drum.
Rotating drums partly filled with granular matter are often used
in chemical engineering for mixing and separation of particles.
Flows in rotating drums recently became a subject of active
research in the physics community. For not too high rotating rates
the flow regime in the drum is separated into an almost solid-body
rotation in the bulk of the drum and a localized fluidized layer
near the free surface (Fig. \ref{fig_drum}). Slowly rotating drums
exhibit oscillations related to the gradual increase of free
surface angle to the static angle of repose and subsequent fast
relaxation to a lower dynamic repose angle via an avalanche.
Transition to steady flow is observed for the higher rotation rate
\cite{Rachenbach:1990}. Scaling of various flow parameters with
the rotation speed (e.g. the width of the fluidized layer etc) and
development of correlations in ``dry'' and ``wet'' granular matter
was recently studied by \textcite{Tegzes:2002,Tegzes:2003}.

\begin{figure}[ptb]
\caption{Schematics of flow structure in the cross-section of
rotating drum, from \textcite{Khakhar:1997} } \label{fig_drum}
\end{figure}

Rotating drums are typically used to study size segregation in
binary mixtures of granular materials. Two types of size
segregation can be distinguished: radial and axial. Radial
segregation is a relatively fast process and occurs after a few
revolutions of the drum. As a result of radial segregation larger
particles are expelled to the periphery and a core of smaller
particles is formed in the bulk
\cite{Metcalfe:1995,Khakhar:1997,Metcalfe:1998,Ottino:2000}, see
Fig. \ref{fig_radial}.

\begin{figure}[ptb]
\caption{Radial size segregation in a rotating drum,  courtesy of
Wolfgang Losert.} \label{fig_radial}
\end{figure}

\begin{figure}[ptb]
\caption{Long rotating drum showing axial size segregation, from
$http://www.physics.utoronto.ca/nonlinear/$}
\label{fig_axial_bands}
\end{figure}

Axial segregation, occurring in the long drums, happens on a much
longer time scale (hundreds of revolutions). As a result of axial
segregation, bands of segregated materials are formed along the
drum axis \cite{Zik:1994,Hill:1994,Hill:1995}, see for
illustration Fig. \ref{fig_axial_bands}. The segregated bands
exhibit slow coarsening behavior. Even more surprisingly, under
certain conditions axial segregation patterns show oscillatory
behavior and travelling waves \cite{Choo:1997,Fiodor:2003}.
Possible mechanisms leading to axial segregation are discussed in
Sec. \ref{sec:segr}.

\subsection{Grains with complex interactions}
\label{subsection:complex}

Novel  collective behaviors emerge when the interactions between
the grains have additional features caused by shape anisotropy,
interstitial fluid, magnetization or electrical charge, etc. In
this situation short-range collisions, the hallmark of
``traditional'' granular systems,   can be augmented by long-range
forces.

Remarkable patterns including multiple rotating vortices of nearly
vertical rods  are observed in the system of vibrated rods by
\textcite{Blair:2003}, see Fig. \ref{fig_blair1}. The rods jump on
their ends slightly tilted and drift
in the direction of the tilt.

Mechanically \cite{Blair:2003b} or electrostatically
\cite{Snezhko:2005} driven magnetic grains exhibit formation of long
chains, isolated rings or interconnecting networks, see Fig.
\ref{fig_snezhko}. In this situation magnetic dipole-dipole
interaction augments hard-core collisions.

Ordered clusters and nontrivial dynamic states were observed by
\textcite{Voth:2002,Thomas:2004} in a small system of particles vibrated
in liquid (Fig. \ref{fig_voth}). It was shown that fluid-mediated interaction between
particles in a vibrating cavity leads to both long-range attraction
and short-range repulsion. A plethora of nontrivial patterns
including rotating vortices, pulsating rings, chains, hexagons etc
was observed by \textcite{Sapozhnikov:2003a} in the system of
conducting particles in dc electric field immersed in poor
electrolyte (Fig. \ref{fig_sap4}). The nontrivial competition between electrostatic
forces and self-induced electro-hydrodynamic flows determines the
structure of emerging pattern.

Granular systems with complex interactions serve as a  natural
bridge to seemingly different systems such as foams, dense
colloids, dusty plasmas, ferrofluids and many others.

\begin{figure}[ptb]
\caption{Select patterns observed in the system of vertically
vibrated rods with the increase of vibration amplitude: a)
nematic-like gas phase; b) moving domains of nearly vertical rods;
c) multiple rotating vortices; d) single vortex, from
\textcite{Blair:2003}.} \label{fig_blair1}
\end{figure}

\begin{figure}[ptb]
\caption{Structures formed in submonolayer of magnetic
microparticles subjected to alternating magnetic field. Select
structures such as rings, compact clusters, and chains are shown
in the top panel. Changes in the pattern morphology with the
increase of magnetic field frequency are illustrated by the three
bottom images, from \textcite{Snezhko:2005}.} \label{fig_snezhko}
\end{figure}

\begin{figure}
\caption{Regular arrangements of particles near the bottom of a
vibrated container filled with water when both attraction and
repulsion are important. All images are taken taken at the
frequency $f=20$ Hz and for different value of dimensionless
acceleration or for different initial conditions: (a) \& (b)
$\Gamma=3$; (c) \& (d) $\Gamma=3.7$ and  $\Gamma$: (e)
$\Gamma=3.9$ and (f) $ \Gamma=3.5$, from \textcite{Voth:2002}.}
\label{fig_voth}
\end{figure}

\begin{figure}
\caption{Representative patterns obtained for different values of
applied field and concentration of ethanol in electrostatically
driven granular system: static clusters (a) and honeycombs (b) and
dynamic vortices (c) and pulsating rings (d),  from
\textcite{Sapozhnikov:2003a}} \label{fig_sap4}
\end{figure}

\section{Main theoretical concepts}

Physics of granular media is a diverse and eclectic field
incorporating many different concepts and ideas, from
hydrodynamics to the theory of glasses. Consequently, many different
theoretical approaches have been proposed to address observed
phenomena.

\subsection{Kinetic theory and hydrodynamics}
\label{subsec:hydrodynamic}

Kinetic theory deals with the  equations for the probability
distributions functions describing the state of granular gas. The
corresponding equations, similar to Boltzmann equations for
rarefied gases,  can be rigorously derived for the dilute gas of
inelastically colliding particles with fixed restitution
coefficient, although certain generalizations are known,
\cite{Goldstein:1995,Jenkins:2002}. Kinetic theory is formulated
in terms of the Boltzmann-Enskog equation for the
probability distribution function $f({\bf v},{\bf r}, t)$ to find the particles
with the velocity ${\bf v}$ at point ${\bf r}$ at time $t$. In the simplest case of
identical frictionless spherical particles  of radius $d$ with fixed restitution coefficient $e$
it assumes the following form
\begin{equation}
\left(\partial_t +({\bf v}_1\cdot \nabla)\right)f(({\bf v}_1,{\bf r}_1,t )
= I[f] \label{bolt}
\end{equation}
with the binary collision integral $I[f]$ in the form
\begin{eqnarray}
I&=&d^2 \int d {\bf v}_2 \int d {\bf n}_{12} \Theta(-{\bf v}_{12} \cdot
{\bf n}_{12})  |{\bf v}_{12} \cdot {\bf n}_{12}| \times \nonumber \\
 &&\left[  \chi  f({\bf v_1}^{\prime \prime}, {\bf r}_1,t ) f({\bf
v_2}^{\prime \prime}, {\bf r}_1-d{\bf n}{_12},t ) \right. \nonumber \\
&& \left.  - f({\bf v}_1,{\bf r}_1,t ) f({\bf v}_2, {\bf r}_1+d{\bf
n}_{12},t)\right]
\end{eqnarray}
where $\chi=1/e^2$, $\Theta$ is
theta-function, and  pre-collision velocities $v_{1,2}$ and
``inverse collision'' velocities $v_{1,2}^{\prime \prime}$ are
related as follows
\begin{equation} {\bf v}_{1,2}^{\prime \prime}  ={\bf
v}_{1,2} \mp \frac{1+e}{2 e }[{\bf n}_{12}({\bf v}_{1}-{\bf
v}_{2})]{\bf n}_{12} \label{collis1}
\end{equation}
(cf. Eq. (\ref{collis})). This equation is derived with the usual
``molecular chaos'' approximation which implies that all
correlations between colliding particles are neglected. One should
keep in mind, however, that in dense granular systems this
approximation can be rather poor due to excluded volume effects
and inelasticity of  collisions introducing velocity correlations
among particles (see, for example, \textcite{Brilliantov:2004}).

Hydrodynamic equations are obtained by truncating the hierarchy of
moment equations obtained from the Boltzmann equation (\ref{bolt}) via
an appropriately modified Chapman-Enskog procedure
(see, e.g., \cite{Jenkins:1985,Brey:1998,Garzo:1999}).
As a result, a set
of continuity equations for mass, momentum and fluctuation kinetic
energy (or ``granular temperature'') is obtained.
However, in
contrast to conventional hydrodynamics, the applicability of
granular hydrodynamics is often questionable because typically  there is
no separation of scale between microscopic and macroscopic motions
\footnote{Except the case of  almost elastic particles with the
restitution coefficient $e \to 1$}, see e.g. \textcite{Tan:1998}.

The mass, momentum and energy conservation equations in granular
hydrodynamics have the form

\begin{eqnarray}
\frac{D{\bf \nu}}{Dt}&=&-\nu \nabla\cdot {\bf u},
\label{mass}\\
\nu\frac{D{\bf u}}{Dt}&=&-\nabla\cdot \bsigma +\nu {\bf g},
\label{momentum}\\
\nu\frac{DT}{Dt}&=&-\bsigma : \dot{\bf \bgamma} - \nabla\cdot {\bf
q}- \varepsilon, \label{energy}
\end{eqnarray}
where $\nu$ is the filling fraction (the density of granular material normalized by the
density of grains), ${\bf u}$ is the velocity field,
$T=(\langle {\bf u}{\bf u}\rangle -\langle {\bf u}\rangle^2) /2$
is the granular temperature, $D/Dt=\partial_t+({\bf
u}\cdot\nabla)$ is the material derivative, ${\bf g}$ is the
gravity acceleration, $\sigma_{\alpha \beta}$ is the stress
tensor,  ${\bf q}$ is the energy flux vector, $\dot\gamma_{\alpha
\beta}=\partial_\alpha u_\beta+\partial_\beta u_\alpha$ is the
strain rate tensor, and $\varepsilon$ is the energy dissipation
rate. Eqs. (\ref{mass})-(\ref{energy}) are structurally similar to
the Navier-Stokes equations for conventional fluids except for the
last term in the equation for granular temperature $\varepsilon$
which accounts for the energy loss due to inelastic collisions.

These three equations have to be supplemented by the constitutive
relations for the stress tensor $\bsigma$, energy flux ${\bf q}$,
and the energy dissipation rate $\varepsilon$. For dilute systems,
a linear relations between stress $ \bsigma$ and strain rate $
\dot  \bgamma$ is obtained,

\begin{eqnarray}
\sigma_{\alpha
\beta}&=&[p+(\mu-\lambda)\mbox{Tr}\dot\bgamma]\delta_{\alpha
\beta}-\mu \dot\bgamma_{\alpha \beta},
\label{c1}\\
{\bf q}&=&-\kappa \nabla T. \label{c2}
\end{eqnarray}

In the kinetic theory of two-dimensional gas of slightly inelastic hard
disks by
\textcite{Jenkins:1985}, these equations are closed with the
following equation of state
\begin{equation}
p=\frac{4\nu T}{\pi d^2}[1+(1+e)G(\nu)], \label{state}
\end{equation}
and the expressions for the shear and bulk viscosities
\begin{eqnarray}
\mu&=&\frac{\nu T^{1/2}}{2\pi^{1/2} d G(\nu)}
\left[1+2G(\nu)+\left(1+\frac{8}{\pi}\right)G(\nu)^2\right],
\label{shear}\\
\lambda&=&\frac{8\nu G(\nu)T^{1/2}}{\pi^{3/2} d}, \label{bulk}
\end{eqnarray}
the thermal conductivity
\begin{equation}
\kappa=\frac{2\nu T^{1/2}}{\pi^{1/2} d G(\nu)}
\left[1+3G(\nu)+\left(\frac{9}{4}+\frac{4}{\pi}\right)G(\nu)^2\right],
\label{kappa}
\end{equation}
and the energy dissipation rate
\begin{equation}
\varepsilon=\frac{16\nu G(\nu)T^{3/2}}{\pi^{3/2} d^3}(1-e^2).
\label{eps1}
\end{equation}
The radial pair distribution function $G(\nu)$ for a dilute 2D gas of
elastic hard disks can be approximated  by the formula \cite{Song:1989}
\begin{equation}
G_{CS}(\nu)=\frac{\nu(1-7\nu/16)}{(1-\nu)^2} \label{gCS}
\end{equation}
(this is a two-dimensional analog of the famous Carnahan-Starling
formula \cite{Carnahan:1969} for elastic spheres). This formula is
expected to work for densities roughly below 0.7. For high density
granular gases, this function has been calculated using free
volume theory by \textcite{Buehler:1951},
\begin{equation}
G_{FV}=\frac{1}{(1+e)\left[(\nu_c/\nu)^{1/2}-1\right]} \label{gFV}
\end{equation}
where $\nu_c\approx 0.82$ is the density of the random close
packing limit.  \textcite{Luding:2001} proposed a global fit
$$
G_{L}=G_{CS}+(1+\exp(-(\nu-\nu_0)/m_0))^{-1})(G_{FV}-G_{CS})
$$
with empirically fitted parameters $\nu_0\approx0.7$ and
$m_0\approx10^{-2}$.  However, even with this extension, the
continuum theory comprised of Eqs.(\ref{mass})-(\ref{eps1}) cannot
describe the force chains which transmit stress via persistent
contacts remaining in the dense granular flows, as well as the
hysteretic transition from solid to static regimes and coexisting solid and
fluid phases.

The granular hydrodynamics is probably the most universal (however
not always the most appropriate) tool for  modelling large-scale
collective behavior in driven granular matter. Granular
hydrodynamics equations in the form
(\ref{mass}),(\ref{momentum}),(\ref{energy}) and their
modifications  are widely used in the engineering community to
describe a variety of large-scale granular flows, especially for
design of gas-fluidized bed reactors \cite{Gidaspow:1994}. In the
physics community granular hydrodynamics is used to understand
various instabilities in relatively small-scale flows, such as
flow past obstacle \cite{Rericha:2002}, convection
\cite{Livne:2002a,Livne:2002b}, floating clusters
\cite{Meerson:2003}, longitudinal rolls
\cite{Forterre:2002,Forterre:2003}, patterns in vibrated layers
\cite{Bougie:2005} and others. However, Eqs.
(\ref{mass})-(\ref{energy}) are often used far beyond their
applicability limits, viz. dilute flows. Consequently, certain
parameters and constitutive relations need to be adjusted
heuristically in order to accommodate observed behavior.  For
example, \textcite{Bougie:2002} had to introduce artificial
non-zero viscosity in Eq. (\ref{momentum}) for $\nu \to 0$ in
order to avoid artificial blowup of the solution. Similarly,
\textcite{Losert:2000} introduced the viscosity diverging as
density approaches the close packed limit as $(\nu-\nu_c)^\beta$
with $\beta\approx 1.75$ being the fitting parameter in order to
describe the structure of dense shear granular flows.

\subsection{Phenomenological models}

A generic  approach to the description of dense granular flows was
suggested by \textcite{Aranson:2001,Aranson:2002b} who proposed to treat
the shear stress mediated fluidization of granular matter as a phase
transition. For this purpose an order parameter characterizing the local
state of granular matter and the corresponding phase field model were
introduced.  According to the model, the order parameter has its own
relaxation dynamics and defines the static and dynamic contributions to
the shear stress tensor. This approach is discussed in more details in
Sec.  \ref{subsubsec:part}.

Another popular approach  is based on the two-phase description of
granular flow, one phase corresponding to rolling grains and the other
phase to static ones. This approach, so-called the BCRE model, was
suggested by \textcite{Bouchaud:1994,Bouchaud:1995} for
description of surface gravity driven flows.  The BCRE model has
direct relation to depth-averaged hydrodynamic equations
(so-called Saint-Venant model) popular in the engineering community.
Note that BCRE and Saint-Venant models can be derived in a certain
limit from the more general order parameter model mentioned
above, for detail see Sec. \ref{subsubsec:two}.

Many pattern-forming systems are often described by generic
amplitude equations such as Ginzburg-Landau or Swift-Hohenberg
equations \cite{Cross:1993,AransonKramer:2002}. This approach
allows to explain many generic features of patterns, however in
any particular system there are peculiarities which need to be
taken into account. This often requires modifications to be
introduced into the generic models. This approach was taken by
\textcite{Tsimring:1997,Aranson:1998,Aranson:1999a,Venkataramani:1998,Crawford:1999}
in order to describe patterns in a vibrated granular layer.
Details of these approaches can be found in Sec.
\ref{subsect:phenom_mod_patt}.
%

In addition, a variety of tools of statistical physics are applied
to diverse phenomena occurring in granular systems.  For example,
celebrated theory of \textcite{Lifsitz:1958} developed for
coarsening phenomena in equilibrium systems was successfully
applied to coarsening of clusters in granular systems
\cite{Aranson:2002}, see Section \ref{subsec:electro}.

\subsection{Molecular dynamics simulations}
\label{subsection:md}

Realistic simulation of granular matter consisting of thousands
of particles remains a challenge for physics and
computer science. Due to simplicity of microscopic interaction
laws (at least for ``dry" and non-cohesive granular matter) and
relatively small number of particles in  granular flows as compared
to atomic and molecular systems, the molecular dynamics
simulations or discrete element models have a potential to address
adequately many phenomena occurring in the granular systems.

There exist three fundamentally different approaches, so-called
soft particles  simulation method;   event driven algorithm and
the contact dynamics  method for rigid particles. For the review
on various molecular dynamics  simulation methods we recommend
\textcite{Rapaport:1995,Luding:2004,Poeschel:2005}.

In the soft particle algorithm, all forces acting on a particle
either from walls or other particles or external forces  are
calculated based on the positions of the particles. Once the
forces are found, the time is advanced by the explicit integration
of the corresponding Newton equations of motion. Various models
are used for calculating normal and tangential contact forces. In
majority of implementations, the normal contact forces are
determined from the particle overlap $\Delta_n$ which is defined
as the difference of the distance between the centers of mass of
two particles and the sum of their radii. The normal force ${\bf
F}_n$ is either proportional to $\Delta_n$ (linear Hookian
contact) or proportional to $\Delta^{3/2}$ (Hertzian contact). In
the spring-dashpot model, additional dissipative force
proportional to the normal component of the relative velocity is
added to model inelasticity of grains.  A variety of approaches
are used to model tangential forces, the most widely accepted of them
being Cundall-Strack algorithm \cite{Cundall:1979}, in which the
tangential contact is modelled by a dissipative linear spring
whose force ${\bf F}_t=-k_t{\bf \Delta}_t -m/2\gamma_t {\bf v}_t$
(here ${\bf \Delta}_t$ is the relative tangential displacement and
${\bf v}_t$ is the relative tangential velocity, $k_t,\gamma_t$
are model constants). It is truncated when its ratio to the normal
force $|{\bf F}_t|/|{\bf F}_n|$ reaches the friction coefficient
$\mu$ according to the Coloumb law. Soft-particles methods are
relatively slow and used mostly for the analysis of dense flows
when generally faster event-driven algorithms are not applicable,
see e.g. \textcite{Silbert:2002a,Silbert:2002b,
Silbert:2002c,Landry:2003,Volfson:2003a,Volfson:2003b}.

In the event-driven  algorithm, the particles are considered
infinitely rigid and move freely  (or driven by macroscopic
external fields) in the intervals between (instantaneous)
collisions.  The algorithm updates velocities and positions of the
two particles involved in a binary collision (in the simplest
frictionless case, according to  Eq. (\ref{collis})), and then
finds the time of the next collision and velocities and positions
of all particles at that time according to Newton's law. Thus, the
time is advanced directly from one collision to the next, and so
variable time step is dictated  by the interval between the
collisions. While event-driven methods are typically faster for
dilute rapid granular flows, they  become  impractical for dense
flows where collisions are very frequent and furthermore particles
develop persistent contacts. As a related numerical problem,
event-driven methods are known to suffer from so-called
``inelastic collapse'' when the number of collisions between
particles diverges in finite time \cite{McNamara:1996}. There are
certain modifications to this method which allow to circumvent
this problem by introducing velocity dependent restitution
coefficient (see, e.g. \textcite{Bizon:1998a}), but still
event-driven methods are mostly applied to rapid granular flows,
see e.g.
\textcite{Ferguson:2004,McNamara:1996,Khain:2004,Nie:2002}.

Contact dynamics is a discrete element method like soft-particles and event-driven ones, with
the equations of motion integrated for each particle. Similarly to
event-driven algorithm and  unlike soft-particles method, particle deformations are
suppressed by considering particles infinitely rigid. The contact dynamics
method considers all contacts occurring within a certain short
time interval as simultaneous, and computes all contact forces by
satisfying simultaneously all kinematic constraints imposed by
impenetrability of the particles and the Coulomb friction law.
Imposing kinematic constraints requires contact forces (constraint
forces) which cannot be calculated from the positions and
velocities of particles alone. The constraint forces are
determined in such a way that constraint-violating accelerations
are compensated. For comprehensive review on the contact dynamics
see \textcite{Brendel:2004}.

Sometimes different molecular dynamics methods are often applied to
the same problem. \textcite{Lois:2005,Staron:2002,Radjai:1998}
applied contact dynamics methods and
\textcite{Silbert:2002a,Silbert:2002b,Volfson:2003a,Volfson:2003b}
used soft-particles technique for  the analysis of instabilities
and constitutive relations in dense granular systems. Patterns in
vibrated layers were studied by event-driven simulations by
\textcite{Bizon:1998a,Moon:2003} and by soft particles molecular
dynamics simulations by \textcite{Prevost:2004,Nie:2000}.

\section{Patterns in sub-monolayers.
Clustering, Coarsening and Phase Transitions}
\label{sec:submon}

\subsection{Clustering in Freely Cooling Gases}
\label{sec:cooling}

Properties of granular  gases are dramatically different from the
properties of molecular gases due to inelasticity of collisions
between the grains. This leads to the emergence of
correlation between colliding particles and violation of the
molecular chaos approximation.  This in turn gives rise to various
pattern-forming instabilities. Perhaps the simplest system
exhibiting nontrivial pattern formation in the context of granular
matter is freely cooling granular gas: isolated system of
inelastically colliding particles. The interest to freely cooling
granular gases was triggered by the discovery of clustering by
\textcite{Goldhirsch:1993}: spontaneously forming dense clusters
emerge as a result of instability of initially homogeneous cooling
state, see Fig. \ref{fig_goldhirsch}. This instability, which can
be traced in many other granular systems, has a very simple
physical interpretation: local increase of the density of granular
gas results in the increase in the number of collisions, and,
therefore, further dissipation of energy and decrease in the
granular temperature. Due to proportionality of pressure to the
temperature, the decrease of temperature will consequently
decrease local pressure, which, in turn, will create a flux of
particles towards this pressure depression, and further increase of
the density. This clustering instability has interesting
counterparts in astrophysics: clustering of self-gravitating gas
\cite{Shandarin:1989} and ``radiative instability'' in optically
thin plasmas  \cite{Meerson:1996} resulting in interstellar dust
condensation.

According to \textcite{Goldhirsch:1993}, the initial stage of
clustering  can be understood in terms of the instability of a
homogeneously cooled state described by the density $\nu$ and
granular temperature $T$. This state is characterized by zero
hydrodynamic velocity $v$, and the temperature evolution follows
from the energy balance equation
\begin{equation}
\partial_t T \sim -T^{3/2}
\label{temp1}
\end{equation}
which results in the Haff's cooling law $T \sim t^{-2}$
\cite{Haff:1983}. However, the uniform cooling state becomes
unstable in large enough systems masking Haff's law. The
discussion of the linear instability conditions can be found e.g.
in \textcite{Babic:1993,Brilliantov:2004}. For the case of
particles with fixed restitution coefficient $e$,  the analysis in
the framework of linearized hydrodynamics equations
(\ref{mass})-(\ref{energy}) yields the critical wavenumber $k^*$
for the clustering instability
\begin{equation}
k^* \sim \sqrt {1-e^2}
\label{kstar}
\end{equation}
As one sees, the length scale of the clustering instability diverges in the limit of elastic particles $e \to 1$.

The clustering instability in a system of grains with constant
restitution coefficient results in the inelastic collapse
discussed in the previous Section.  Whereas the onset of
clustering can be well-understood in the framework of granular
hydrodynamics (see, e.g.
\textcite{Babic:1993,Hill:2003,Goldhirsch:2003,Brilliantov:2004}),
certain subtle features (e.g. scaling exponents for temperature)
are only assessed within molecular dynamics simulation because the
hydrodynamic description often breaks in dense cold clusters. One
recent theoretical approach to the description of the late stages
of clustering instability consists in introducing additional
regularization into the  hydrodynamic description due to the
finite size of particles \cite{Nie:2002,Efrati:2005}.

\textcite{Nie:2002} argued that cluster formation and coalescence
in freely cooling granular gases can be heuristically described by
the Burgers equation for hydrodynamic velocity $v$ with random
initial conditions:
\begin{equation}
\partial_t v+ v \nabla v = \mu_0 \nabla^2 v
\label{burgers}
\end{equation}
where $\mu_0$ is effective viscosity (which is
different from the shear viscosity in hydrodynamic equations). In
this context clustering is associated with the formation of shocks
in the Burgers equation. Perhaps not surprising, a very similar
approach was applied for description of the gas of ``sticky"
particles for the description of the large-scale matter formation
in the Universe \cite{Gurbatov:1985,Shandarin:1989}.

\textcite{Meerson:2005} conducted molecular dynamics simulations
of  the clustering instability of a freely cooling dilute
inelastic gas in a quasi-one-dimensional setting.
This problem was also examined in the framework of
granular hydrodynamics by \textcite{Efrati:2005}.
It was observed
that, as the gas cools, stresses become negligibly small,
and the gas flows only by inertia.
Hydrodynamic description reveals  a finite-time singularity,
as the velocity gradient and the gas density diverge at some location.
The molecular dynamics studies show that finite-time
singularities, intrinsic in such flows, are arrested only when
close-packed clusters are formed. It was confirmed that the
late-time dynamics and coarsening behavior are describable by the
Burgers equation (\ref{burgers}) with vanishing viscosity $\mu_0$.
Correspondingly, the average cluster mass grows as $t^{2/3}$ and
the average velocity decreases as $t^{-1/3}$. Due to the
clustering long-term temperature evolution is $T \sim t^{-2/3} $
which is   different from Haff's law $T\sim t^{-2}$ derived for
the spatially-homogeneous cooling.
\textcite{Efrati:2005} argue that flow by inertia represents a
generic intermediate asymptotic of unstable free cooling of dilute
granular gases consistent with the Burgers equation
(\ref{burgers}) description of one-dimensional gas of ``sticky
particles'' suggested by \textcite{Nie:2002}.

While there is a qualitative similarity between Burgers shocks and
clusters in granular materials at least in one dimension, the
applicability of the Burgers equation for the description of
granular media is still an open question, especially in two and
three dimensions. The main problem is that the Burgers equation
can be derived from the hydrodynamic equations only in one
dimensional situation, in two and three dimensions the Burgers
equation assumes {\it zero vorticity}, which possibly
oversimplifies the problem and may miss important physics. In
fact, molecular dynamics simulations illustrate the development of
large-scale vortex flows in the course of clustering instability
\cite{Catuto:2004,vanNoije:2000}.

Finally, \textcite{Das:2003} proposed  a phenomenological
description of the long-term clusters evolution in granular gases.
Using the analogy between clustering in granular gases and
phase-ordering dynamics in two-component mixtures,
\textcite{Das:2003} postulated  generalized Cahn-Hilliard
equations for the evolution of density $\nu$ and complex velocity
$\psi=v_x+i v_y$  (see e.g. \cite{Bray:1994})
\begin{eqnarray}
\partial_t \nu &= &(-\nabla^2)^m \left[ \nu - \nu^3 + \nabla^2 \nu
\right] \label{den1} \\
\partial_t \psi &= &(-\nabla^2)^m \left[ \psi - |\psi|^2 \psi + \nabla^2
\psi \right] \label{vel1}
\end{eqnarray}
with $m\to 0+$ which characterize globally-conserved dynamics of
$\nu$ and $\psi$ similar to that considered in Sec.
\ref{sec:coars2}. \textcite{Das:2003} argue that this choice is
most appropriate due to the non-diffusive character of particles
motion and is consistent with the observed morphology of clusters.
While  it might be very challenging to derive Eqs.
(\ref{den1}),(\ref{vel1}) from the first principles or to deduce
them  from hydrodynamic equations, the connection to
phase-ordering dynamics is certainly deserves further
investigation.

\subsection{Patterns in Driven Granular Gases}
Discovery of the clustering instability stimulated a large number of
experimental and theoretical studies, even experiments in low
gravity conditions \cite{Falcon:1999}. Since ``freely cooling
granular gas" is difficult to implement in the laboratory, most
experiments were performed in the situation when the energy is
injected in the granular system in one or another way.
\textcite{Kudrolli:1997}  studied two-dimensional granular
assemblies interacting with a horizontally vibrating (or ``hot'')
wall. In agreement with granular hydrodynamics, maximum gas
density occurs {\it opposite} to the vibrating wall, see Fig.
\ref{fig_gollub}.  The experimental density distributions are
consistent with the modified hydrodynamic approach proposed by
\textcite{Grossman:1997}.
\textcite{Livne:2002a,Livne:2002b,Khain:2002,Khain:2004a}, studied
the dynamics of granular gases interacting with a hot wall
analytically using granular hydrodynamic theory for rigid disks in
the formulation of \textcite{Jenkins:1985} and predicted a novel
phase-separation or van der Waals-type instability of the
one-dimensional density distribution. This instability, reproduced
later by molecular dynamics simulations \cite{Argentina:2002}  is different
from the usual convection instability as it occurs without gravity
and is driven by the coarsening mechanism. Simulations indicated
a profound role of fluctuations. One may expect that noise
amplification near the instability thresholds in granular systems
will be very important due to non-macroscopic number of grains. In
the context of phase-separation instability
\textcite{Meerson:2004} raised the non-trivial question of the origin
of giant fluctuations and break-down of hydrodynamic description
in granular systems near the threshold of instability (see also
\cite{Goldman:2004,Bougie:2005} on the effect of fluctuations in
multilayers). Remarkably, for the granular gas confined between
two oscillating walls \textcite{Khain:2004} predicted on the basis
of event-driven simulations a novel oscillatory instability for the
position of the dense cluster. These predictions, however, have not
yet been confirmed experimentally, most likely due to the relatively
small aspect ratio of available experimental cells.

\begin{figure}[ptb]
\caption{Sample image showing dense cold cluster formed opposite
the driving wall (at the bottom), total number of particles 1860,
from  \textcite{Kudrolli:1997}.} \label{fig_gollub}
\end{figure}

\textcite{Olafsen:1998} pioneered experiments with sub-monolayers
 of particles subject to vertical vibration\footnote{Sub-monolayer
implies less than 100\% percent coverage by particles of the
bottom plate.}. Their studies revealed a surprising phenomenon:
formation of a dense closely-packed cluster co-existing with
dilute granular gas, see Fig. \ref{fig_olafsen}. The phenomenon
bears a strong resemblance to the first-order solid/liquid phase
transition in equilibrium systems. Similar experiments by
\textcite{Losert:1999} discovered propagating fronts between
gas-like and solid-like phases in vertically vibrated
sub-monolayers. Such fronts are expected in extended  systems in
the vicinity of the first order phase transition, e.g.
solidification fronts in  supercooled liquids.
\textcite{Prevost:2004} performed
 experiments with vibrated granular gas confined between two
 plates. Qualitatively similar phase coexistence was found.
The cluster formation in vibro-fluidized sub-monolayers shares
many common features with processes in freely cooling granular
gases because it is also caused by the energy dissipation due to
inelasticity of collisions. However, there is a significant
difference: the instability described in Subsec. \ref{sec:cooling}
is insufficient to explain the phase separation. A
very important additional factor is  {\it bistability} and
co-existence of states due to the nontrivial density dependence of
the transfer rate of particle's vertical to horizontal momentum.
Particles in a dense closed-packed cluster likely obtain less
horizontal momentum than in a moderately dilute gas because in the
former particle vibrations are constrained to the vertical plane
by interaction with neighbors. In turn, in a very dilute gas the
vertical to horizontal momentum transfer is also inhibited due to
lack of particle collisions. Another factor here is that vibration
is not fully equivalent to the interaction with a heat bath. It is
well known that even a single particle interacting with a
periodically vibrating plate exhibits coexistence  of dynamic and
static states \cite{Losert:1999}.

There were several simulation studies of  clustering and phase
coexistence in vibrated granular submonolayers.
\textcite{Nie:2000,Prevost:2004} reproduced certain features of
cluster formation and two-phase co-existence by means of
large-scale three-dimensional molecular dynamics simulations.
Since realistic three-dimensional simulations are still expensive
and extremely time-consuming, simplified modelling of the effect
of a vibrating wall by a certain multiplicative random forcing on
individual particles was employed by \textcite{Cafiero:2000}.
While the multiplicative random forcing is an interesting
theoretical idea, it has to be used with caution as it is not
guaranteed to reproduce subtle details of particle dynamics,
especially the sensitive dependence of the vertical to horizontal
momentum transfer as the function of the density.

\subsection{Coarsening of clusters}
\label{sec:coars1}

One of the most intriguing questions in the context of phase
coexistence in vibrofluidized granular sub-monolayers is a
possibility of Ostwald-type ripening and coarsening of clusters
similar to that observed in equilibrium systems
\cite{Lifsitz:1958,Lifsitz:1961}. In particular, the scaling law
for the number of macroscopic clusters is of special interest
because it gives  a deep insight into the similarity between
equilibrium thermodynamic systems and non-equilibrium granular
systems. The experiments
\cite{Olafsen:1998,Losert:1999,Prevost:2004,Sapozhnikov:2003}
demonstrated emergence and growth of multiple clusters but did not
have sufficient aspect ratio to address the problem of coarsening
in a quantitative way.

Nevertheless, as it was suggested by \textcite{Aranson:2000ed},
statistical information on out-of-equilibrium Ostwald ripening can
be obtained in a different granular system:
electrostatically driven granular media. This system permits one to
operate with extremely small particles and obtain a very large
number of macroscopic clusters. In this system the number of
clusters $N$ decays with time as $N \sim 1/t$. This law
is consistent with interface-controlled Ostwald ripening in two
dimensions, see \cite{Wagner:1961}. Whereas mechanisms of energy
injections are different, both vibrofluidized and
electrostatically-driven systems show similar behavior:
macroscopic phase separation, coarsening, transition from two- to
three-dimensional cluster growth, etc \cite{Sapozhnikov:2003}. In
\textcite{Aranson:2002} the theoretical description of granular
coarsening was developed in application to the electrostatically
driven grains, however we postpone the description of this theory
to Sec. \ref{subsec:electro}. We anticipate that a theory
similar to that formulated in \textcite{Aranson:2002} can be
applicable to mechanically fluidized granular materials as well.
The main difference there is the physical mechanism of energy
injection which will possibly affect the specific form of the
conversion rate function $\phi$ in Eq. (\ref{gle1}) in Sec.
\ref{subsec:electro}.

\section{Surface waves and patterns in vibrated multilayers of granular
materials} \label{sec:multil}

\subsection{Chladni patterns and heaping}
Driven granular systems often manifest collective fluid-like
behavior: shear flows, convection, surface waves, and pattern
formation (see e.g. \textcite{Jaeger:1996}). Surprisingly, even very
thin (less than ten) layers of sand under excitation exhibit
pattern formation which is quite similar (however with some
important differences) to the corresponding patterns in fluids.  One
of the most fascinating examples of  these collective dynamics is
the appearance of long-range coherent patterns and localized
excitations in vertically-vibrated thin granular layers.

Experimental studies of vibrated layers of sand have a long and
illustrious history, beginning from the seminal works by
\textcite{Chladni:1787} and \textcite{Faraday:1831} in which they
used a violin bow and a membrane to excite vertical vibrations in
a thin layer of grains. The main effect observed in those early
papers, was ``heaping'' of granular matter in mounds near the
nodal lines of the membrane oscillations. This behavior was
immediately (and correctly) attributed to the ``acoustic
streaming'', or nonlinear detection of the nonuniform excitation
of grains by membrane modes. One puzzling result by Chladni was
that a very thin powder would collect at the anti-nodal regions
where the amplitude of vibrations is maximal. As Faraday
demonstrated by evacuating the container, this phenomenology is
caused by the role of air permeating the grains in motion.
Evidently, the interstitial gas becomes important as the terminal
velocity of a free fall  $v_t=\nu g d^2/18\mu$ becomes of the
order of the plate velocity, and this condition is fulfilled for
$10-20$ $  \mu m$ particles on a plate vibrating with frequency 50
Hz and acceleration amplitude $g$.

In subsequent years the focus of attention was diverted from
dynamical properties of thin layers of vibrated sand, and only in
the last third of the 20th century physicists returned to this old
problem equipped with new experimental capabilities. The dawn of
the new era was marked by the studies of heaping by
\textcite{Jenny:1964}. In subsequent papers
\cite{Walker:1982,Dinkelacker:1987,Evesque:1989,Douady:1989,Laroche:1989},
more research has been performed of heaping with and without
interstitial gas, with somewhat controversial conclusions on the
necessity of ambient gas for heaping (see, e.g.
\cite{Evesque:1990}). Eventually, after more careful analysis
\textcite{Pak:1995} concluded that heaping indeed disappears as
the pressure of the ambient gas tends to zero or the particle size
increases. This agreed with numerical molecular dynamics simulations
\cite{Taguchi:1992,Gallas:1992a,Gallas:1992b,Gallas:1992c,Gallas:1993,Luding:1994}
which showed no heaping without interstitial gas effects.  Recent
studies of deep layers ($50<N<200$) of small particles ($10<d<200\mu m$)
by \textcite{Falcon:1999a,Duran:2000,Duran:2001} showed a number of
interesting patterns and novel instabilities caused by interstitial air. In
particular, \textcite{Duran:2001} observed formation of isolated
droplets of grains after periodic taping similar to the Rayleigh-Taylor instability in
ordinary fluids.

\textcite{Jia:1999} proposed a simple model for heap formation
which is motivated by these experiments. In a discrete lattice
version of the model, the decrease in local density due to
vibrations is modelled by the random creation of empty sites in
the bulk. The bulk flow is simulated by the dynamics of empty
sites, while the surface flow is modelled by rules similar to the
sandpile model (see Sec. \ref{sec:SOC}). This model
reproduced both convection inside the powder and the heap formation
for sufficiently large probability of empty site formation (which
mimics the magnitude of vibration). \textcite{Jia:1999} also
proposed the continuum model which has a simple form of a
nonlinear reaction-diffusion equation,  for the local height of
the sandpile
\begin{equation}
\partial_t h=D\nabla^2h +\Omega h - \beta h^2.
\label{eq_jia}
\end{equation}
However this model is perhaps too
generic and lacks the specific physics of the heaping process.

\subsection{Standing wave patterns}
While heaping may or may not appear depending on the gas pressure
and the particle properties at small vertical acceleration, at
higher vertical acceleration patterns of standing waves emerge in
thin layers.  They were first reported by
\textcite{Fauve:1989,Douady:1989} in a quasi two-dimensional
geometry. These waves oscillated at the half of the driving
frequency, which indicates the sub-harmonic resonance
characteristic for parametric instability.  This first observation
spurred a number of experimental studies of standing waves in thin
granular layers in two  and three dimensional  geometries
\cite{Melo:1994,Melo:1995,Umbanhowar:1996,Clement:1996,Aranson:1999b,Mujica:1998}.
Importantly, these studies were performed in evacuated containers,
which allowed to obtain reproducible results not contaminated by
heaping. Fig. \ref{fig_vibr} shows a variety of regular patterns
observed in vibrated granular layers under vibration
\cite{Melo:1994}. As a result of these studies, the emerging
picture of pattern formation appears as follows.

The particular pattern is determined by the interplay between
driving frequency $f$ and acceleration of the container $\Gamma= 4
\pi^2 {\cal A }  f^2/g$ (${\cal A} $ is the amplitude of
oscillations, $g$ is the gravity acceleration)
\cite{Melo:1994,Melo:1995}. The layer of grains remains flat for
$\Gamma<2.4$ more-less independent of driving frequency. At higher
$\Gamma$ patterns of standing waves emerge. At small frequencies
$f< f^*$ (for experimental conditions of \textcite{Melo:1995}
, $f^*\approx$ 45 Hz) the transition is subcritical,
leading to the formation of square wave patterns, see
Fig.\ref{fig_vibr}b. For  higher frequencies $f >f^* $ the selected
pattern is quasi-one-dimensional stripes (Fig. \ref{fig_vibr}a), and the transition
becomes supercritical. In the intermediate  region $f\sim f^*$,
localized excitations ({\it oscillons},
Fig.\ref{fig_swinney_oscillon}) and various bound states of oscillons
(Fig.\ref{fig_vibr}f) were observed within the hysteretic region
of the parameter plane. Both squares and stripes, as well as
oscillons, oscillate at the half of the driving frequency, which
indicates the parametric mechanism of their excitation. The
wavelength of the cellular patterns near the onset scales linearly with the
depth of the layer and diminishes with the frequency of vibration
\cite{Umbanhowar:2000}. The frequency corresponding to the
strip-square transition was shown to depend on the particle
diameter $d$ as $d^{-1/2}$. This scaling suggests that the
transition is controlled by the relative magnitude of the energy
influx from the vibrating plate $\propto f^2$ and the
gravitational dilation energy $\propto gd$. At higher acceleration
($\Gamma>4 $), stripes and squares become unstable, and hexagons
appear instead (Fig. \ref{fig_vibr}c). Further increase of
acceleration at $\Gamma \approx 4.5$ converts hexagons into a
domain-like structure of flat layers oscillating with frequency
$f/2$ with opposite phases. Depending on parameters, interfaces
separating flat domains, are either smooth or ``decorated'' by
periodic undulations (Fig. \ref{fig_vibr}e). For $\Gamma >5.7$
various quarter-harmonic patterns emerge. The complete phase diagram of
different regimes observed in a three-dimensional  container is
shown in Fig. \ref{fig_swinney_ph_diag}. For even higher
acceleration ($\Gamma>7$) the experiments reveal surprising phase
bubbles and spatio-temporal chaos oscillating approximately at
one fourth the driving frequency \cite{Moon:2002}.

\begin{figure}[ptb]
\caption{Phase diagram of various regimes in vibrated granular
layers, from \textcite{Melo:1995}.} \label{fig_swinney_ph_diag}
\end{figure}

Subsequent investigations revealed that periodic patterns share
many features with convective rolls in Rayleigh-B\'{e}nard
convection, for example skew-varicose and cross-roll instabilities
\cite{Bruyn:1998}.

\subsection{Simulations of vibrated granular layers}

The general understanding of the standing wave patterns in thin
granular layers can be gained by the analogy with ordinary fluids.
The Faraday instability in fluids and corresponding pattern
selection problems have been studied theoretically and numerically
in great detail (see e.g. \textcite{Zhang:1997}). The primary
mechanism of instability is the parametric resonance between the
spatially uniform periodic driving at frequency $f$ and two
counter-propagating gravity waves at frequency $f/2$. However,
this instability in ordinary fluids leads to a supercritical
bifurcation and square wave patterns near offset, and as a whole
the corresponding phase diagram lacks the richness of the granular
system. Of course this can be explained by the fact that there are
many qualitative differences between granular matter and
fluids, such as presence of strong dissipation, friction and
the absence of surface tension in the former. Interestingly,
localized oscillon-type objects were subsequently observed in vertically
vibrated layers of non-Newtonian fluid \cite{Lioubashevski:1999},
and stipe patterns were observed in highly viscous fluid
\cite{Kiyashko:1996}. The theoretical understanding of the pattern
formation in a vibrated granular system presents a challenge,
since unlike fluid dynamics there is no universal theoretical
description of dense granular flows analogous to the Navier-Stokes
equations. In the absence of this common base, theoretical and
computational efforts in describing these patterns followed
several different directions. \textcite{Aoki:1996} were first to
perform molecular dynamics simulations of patterns in the vibrated
granular layer. They concluded that grain-grain friction is
necessary for pattern formation in this system.  However, as noted
by \textcite{Bizon:1997}, this conclusion is a direct consequence
of the fact that the algorithm of \textcite{Aoki:1996}, which is
based on the Lennard-Jones interaction potential and velocity
dependent dissipation, leads to the restitution coefficient of
particles approaching unity for large collision speeds rather than
decreasing according to experiments.

\textcite{Bizon:1998a,Bizon:1998b} performed
event-driven simulations of colliding grains on a vibrated plate assuming
constant restitution
(see also \textcite{Luding:1996} for earlier two-dimensional event-driven simulations).
It was demonstrated that even without friction, patterns do form
in the system, however only supercritical bifurcation to stripes is
observed. It turned out that friction is necessary to produce
other patterns observed in experiments, such as squares and $f/4$
hexagons. Simulations with frictional particles reproduced
the majority of patterns observed in experiments and many features of
the bifurcation diagram (with the important exception of the
oscillons). \textcite{Bizon:1998a}  set out to match an experimental
cell and a numerical system,
maintaining exactly the same size container and sizes and the
number of particles. After fitting only two parameters of the
numerical model, \textcite{Bizon:1998a} were able to find a very
close quantitative agreement between various patterns in the
experimental cell and patterns in simulations throughout the
parameter space of the experiment (frequency of driving, amplitude
of acceleration, thickness of the layer), see Fig.
\ref{fig_swinney_comp_exp_sim}.

\begin{figure}[ptb]
\caption{Comparison between subharmonic patterns in experiment
(left) and three dimensional  molecular dynamics simulations
(right) of 30000 particles in a square vibrated container for
different frequencies and amplitudes of vibration, from
\textcite{Bizon:1998a}.} \label{fig_swinney_comp_exp_sim}
\end{figure}

\textcite{Shinbrot:1997} proposed a model which combined ideas
from molecular dynamics and continuum modelling. Specifically, the
model ignored vertical component of particle motion and assumed
that impact with the plate adds certain randomizing horizontal
velocity to the individual particles. The magnitude of the random
component being added at each impact served as a measure of impact
strength. After the impact particles were allowed to travel freely
in the horizontal plane for a certain fraction of a period after
which they inelastically collide with each other (a particle
acquires momentum averaged over all particles in its
neighborhood). This model did reproduce a variety of patterns seen
in experiments (stripes, squares, and hexagons) for various values
of control parameters (frequency of driving and impact strength),
however it did not describe some of the experimental phenomenology
(localized objects as well as interfaces), besides it also
produced a number of intricate patterns not seen in experiments.

\subsection{Continuum theories}
\label{subsect:phenom_mod_patt}
The first continuum models of pattern formation in vibrating sand
were purely phenomenological. In the spirit of weakly-nonlinear
perturbation theories \textcite{Tsimring:1997} introduced the
complex amplitude $\psi(x,y,t)$ of sub-harmonic oscillations of
the layer surface, $h=\psi\exp(i\pi f t)+c.c.$. The equation for
this function on the symmetry grounds in the lowest order
was written as
\begin{equation}
\partial_t\psi=\gamma\psi^*-(1-i\omega)\psi+(1+ib)\nabla^2\psi
-|\psi|^2\psi-\nu\psi . \label{eq0a}
\end{equation}
Here $\gamma$ is the normalized amplitude of forcing at the
driving frequency $f$.  The linear terms in  Eq. (\ref{eq0a})  can
be obtained from the complex growth rate for infinitesimal
periodic layer perturbations $h \sim  \exp[ \Lambda(k) t + i k
x]$. Expanding $\Lambda(k)$ for small $k$, and keeping only two
leading terms in the expansion $\Lambda (k) = -\Lambda_0
-\Lambda_1 k^2$ gives rise to the linear terms in Eq. (\ref{eq0a}),
where $b = Im \Lambda _1/ Re \Lambda_1$ characterizes ratio of
dispersion to diffusion and parameter $\omega = -(Im \Lambda_0 + \pi f )
/Re \Lambda _0 $, characterizes the frequency of the driving.

The only difference between this equation and the Ginzburg-Landau
equation for the parametric instability \cite{Coullet:1990} is the
coupling of the complex amplitude $\psi$ to the ``slow mode''
$\nu$ which characterizes local dissipation in the granular layer
($\nu$ can be interpreted as coarse-grained layer's number
density). This slow mode obeys its own dynamical equation
\begin{eqnarray}
\partial_t\nu&=&  \alpha
\nabla\cdot(\nu\nabla|\psi|^2)+\beta\nabla^2\nu .\label{eq0b}
\end{eqnarray}

This equation describes re-distribution of the averaged thickness
due to the diffusive flux $\propto -\nabla \nu$, and an
additional flux $\propto -\nu\nabla |\psi|^2$ is caused by the
spatially nonuniform vibrations of the granular material. This
coupled model was used by \textcite{Tsimring:1997,Aranson:1998} to
describe the pattern selection near the threshold of the primary
bifurcation. The phase diagram of various patterns found in this
model is shown in Fig. \ref{fig_phase_diag_theory}. At small
$\alpha \langle\nu\rangle\beta^{-1}$ (which corresponds to low frequencies
and thick layers), the primary bifurcation is subcritical and
leads to the emergence of square patterns. For higher frequencies
and/or thinner layers, transition is supercritical and leads to
roll patterns. At intermediate frequencies stable localized
solutions of Eqs.(\ref{eq0a}),(\ref{eq0b}) corresponding to
isolated {\em oscillons} and a variety of bound states were found
in agreement with experiment. The mechanism of oscillon
stabilization is related to the oscillatory asymptotic behavior of
the tails of the oscillon (see Fig. \ref{fig_oscillon_theory}),
since this underlying periodic structure provides pinning for the
circular front forming the oscillon. Without such pinning, the
oscillon solution could only exist at a certain unique value of a
control parameter (e.g. $\gamma$), and would either collapse or
expand otherwise.

\begin{figure}[ptb]
\caption{Phase diagram showing primary stable patterns derived
from Eqs. (\protect\ref{eq0a}),(\protect\ref{eq0b}). Points
indicate stable oscillons obtained by numerical solution of Eqs.
(\protect\ref{eq0a}),(\protect\ref{eq0b}), $\eta=\alpha/\beta$,
$\mu= \langle \nu \rangle$ is average density, and
$\epsilon\sim\gamma-\gamma_c$ is supercriticality parameter,  from
\textcite{Tsimring:1997}.} \label{fig_phase_diag_theory}
\end{figure}

\begin{figure}[ptb]
\caption{ Radially-symmetric oscillon solution of
Eqs.(\ref{eq0a}),(\ref{eq0b}) for
$\gamma=1.8,\mu=0.567,b=2,\omega=\alpha=1,\eta=5/\gamma$, from
\textcite{Tsimring:1997}.} \label{fig_oscillon_theory}
\end{figure}

Let us note that stable localized solutions somewhat resembling
oscillons have recently been found in the nonlinear
Schr\"{o}dinger equation with additional linear dissipation and
parametric driving \cite{Barashenkov:2002}.

Phenomenological model (\ref{eq0a}),(\ref{eq0b}) also provides a
good description of patterns away from the primary bifurcation -
hexagons and interfaces \cite{Aranson:1999a}. In high-frequency
limit the slow mode dynamics can be neglected ($\nu$ becomes
enslaved by $\psi$), and the dynamics can me described by a single
parametric Ginzburg-Landau equation (\ref{eq0a}).

It is convenient to shift  the phase of the complex order
parameter via $\tilde{\psi}=\psi\exp(i\phi )$ with $\sin
2\phi=\omega/\gamma$. The equations  for real and imaginary part
$\tilde \psi = A + i B$ are:
\begin{eqnarray}
\partial_t A &=&(s -1) A -  2  \omega B - ( A^2+ B^2) A +\nabla^2 ( A- b
B),
\nonumber \\
\partial_t B  &=& -  (s + 1)  B - ( A^2+ B^2) B +\nabla^2 ( B + b  A ),
\label{eq2b}
\end{eqnarray}
where $s^2 = \gamma^2 -\omega^2$. At $s<1$, Eqs. (\ref{eq2b}) has
only one trivial uniform state $A=0,\ B=0$, At $s>1$, two new
uniform states  appear, $A=\pm A_0, B=0, A_0  = \sqrt{s-1}$. The
onset of these states corresponds to the period doubling of the
layer flights sequence,  observed in experiments
\cite{Melo:1994,Melo:1995} and predicted by the simple inelastic
ball model \cite{Melo:1994,Melo:1995,Metha:1990}. Signs $\pm$
reflect two relative phases of layer flights with respect to
container vibrations.

Weakly-nonlinear analysis reveals that the uniform states $\pm
A_0$ lose their stability with respect to finite-wavenumber
perturbations at $s<s_c$, and the nonlinear interaction of growing
modes leads to hexagonal patterns.  The reason for this is that
the non-zero base state $A=\pm A_0$ lacks the up-down symmetry
$\psi\to-\psi$ and the corresponding amplitude equations contains
quadratic terms which are known to favor hexagons close to onset
(see, e.g. \textcite{Cross:1993}). In the regime when the uniform
states $A=\pm A_0,B=0$ are stable, there is an interface solution
connecting these two asymptotic states. This interface may exhibit
transversal instability which leads to decorated interfaces (see
experimental Fig. \ref{fig_vibr}e). Due to symmetry, the
interfaces are immobile, however breaking the symmetry of driving
can lead to interface motion. This symmetry breaking can be
achieved by additional subharmonic driving at frequency $f/2$. The
interface will move depending on the relative phases of $f$ and
$f/2$ harmonics of driving. This interface drift was predicted in
\cite{Aranson:1999a} and observed in the subsequent work
\cite{Aranson:1999b}. As it was noted by \textcite{Aranson:1999b}
(see also later work by \textcite{Moon:2003}), moving interfaces
can be used to separate granular material of different sizes. The
stability and transition between flat and decorated interfaces was
studied theoretically and experimentally by \textcite{Blair:2000}.
It was shown that non-local effects are responsible for the
saturation of transverse instability of interfaces. Moreover, new
localized solutions (``superoscillons'') were found for large
accelerations. In contrast with conventional oscillons existing on
the flat background oscillating with driving frequency $f$, i.e.
in our notation $\psi=0$, the superoscillons exist on the
background of the flat period-doubled solution $\psi \ne 0$.

Another description of the primary pattern-forming bifurcation was
done by \textcite{Crawford:1999} in the framework of the
generalized Swift-Hohenberg equation
\begin{eqnarray}
\partial_t\psi & = & R\psi-(\partial_x^2+1)^2\psi+b\psi^3-c\psi^5
+ \varepsilon\nabla\cdot[(\nabla\psi)^3]  \nonumber \\
&-&\beta_1\psi(\nabla\psi)^2-\beta_2\psi^2\nabla^2\psi.
\label{SwiftHoh}
\end{eqnarray}
Here the (real) function $\psi$ characterizes the amplitude of the
oscillating solution, so implicitly it is assumed that the whole
pattern always oscillates in phase.  Terms proportional to
$\varepsilon$ have been added to the standard
Swift-Hohenberg equation first introduced for description of
convective rolls (see, e.g.  \textcite{Cross:1993}) since they are
known to favor square patterns, and extended fifth-order local
nonlinearity allowed to simulate subcritical bifurcation for
$R<0$. This equation also describes both square and stripe
patterns depending on the magnitude of $\varepsilon$ and for
negative $R$ has a stable oscillon-type solution.

Even more generic approach was taken by
\textcite{Venkataramani:1998,Venkataramani:2001} who argued that
the spatio-temporal dynamics of patterns generated by
parametric forcing can be understood in the framework of a
discrete-time, continuous space system
which locally exhibits a sequence of period-doubling bifurcations
and whose spatial coupling operator selects a certain spatial
scale. In particular they studied the discrete-time system
\begin{equation}
\xi_{n+1}({\bf x})= {\cal L}  [ M(\xi_{n}({\bf x}))]
\label{ven_ott}
\end{equation}
where local mapping $M(\xi)$ is described by a Gaussian map
$$M(\xi)=\tilde r\exp[-(\xi-1)^2/2]$$ and the linear spatial operator
$\cal{L}$ has an azimuthally symmetric Fourier transform
$$
f(k)=\mbox{sign}[k_c^2-k^2]\exp[k^2(1-k^2/2k_0^2))/2].
$$
Here $k$ is the wavenumber, $k_c, k_0$ are two inverse length
scales characterizing the spatial coupling, and $\tilde r$
describes the amplitude of forcing. While this choice of the
spatial operator appears rather arbitrary, it leads to a phase
diagram on the plane $(k_c/k_0,r)$ which is similar to the
experimental one.

Several authors \cite{Cerda:1997,Eggers:1999,Park:2002} attempted
to develop a quasi two-dimensional fluid-dynamics-like continuum
description of the vibrated sand patterns. These models deal with
mass and momentum conservation equations which are augmented by
specific constitutive relations for the mass flux and
pressure. \textcite{Cerda:1997} assumed that during impact
particles acquire horizontal velocities proportional to the
gradient of local thickness, then during the flight that move
freely with these velocities and redistribute mass, and  during
the remainder of the cycle the layer diffusively relaxes on the
plate. The authors found that a flat layer is unstable with
respect to square pattern formation, however the transition is
supercritical. In order to account for the subcritical character
of the primary bifurcation to square patterns, the authors
postulated the existence of a certain critical slope (related to
the repose angle) below which the free flight initiated by the
impact does not occur. They also observed the existence of
localized excitations (oscillons and bound states), however they
appeared only as transients in the model. \textcite{Park:2002}
generalized this model by explicitly writing the momentum
conservation equation and introducing the equation of state for
the hydrodynamic pressure which is
proportional to the square of the velocity divergence. This effect
provides saturation of the free-flight focusing instability and
leads to a squares-to-stripes transition at higher frequencies
which was missing in the original model \cite{Cerda:1997}.  By
introducing multiple free-flight times and contact times
\textcite{Park:2002} were also able to reproduce hexagonal
patterns and superlattices.

Full three-dimensional continuum simulations based on the granular
hydrodynamics equations
(\ref{mass}),(\ref{momentum}),(\ref{energy}) were performed by
\textcite{Bougie:2005}. Quantitative agreement was found between
this description and event-driven molecular dynamics simulations
and experiments in terms of the wavelength dependence on the
vibration frequency (Fig.\ref{fig_swinney_disp_curve}) although
the authors had to introduce a certain regularization procedure in
the hydrodynamic equations in order to avoid artificial numerical
instabilities for $\nu \to 0$. Since standard granular
hydrodynamics does not take into account friction among particles,
the simulations only yielded stripe pattern, in agreement with
earlier molecular dynamics simulations. Furthermore, the authors
found a small but systematic difference ($\sim$10\%) between the
critical value of plate acceleration in fluid-dynamical and
molecular dynamics simulations which could be attributed to the
role of fluctuations near the onset. Proper account of
inter-particle friction and fluctuations within the full
hydrodynamics description still remains an open problem (see more
on that in Section \ref{sec:dense}).

\begin{figure}
\caption{ Dispersion relation for stipes near the onset according
to continuum granular hydrodynamics equations and molecular
dynamics  simulations compared with experimental data, from
\textcite{Bougie:2005}.} \label{fig_swinney_disp_curve}
\end{figure}

Fluctuations are expected to play a significantly greater role in
granular hydrodynamics than in usual fluids, because the total
number of particles involved in the dynamics per characteristic
spatial scale of the problem is many orders of magnitude smaller
than the Avogadro number. The apparatus of
fluctuating hydrodynamics which was developed in particular for
description of transition to rolls in Rayleigh-B\'{e}nard
convection \cite{Swift:1977}, has been recently applied to
the granular patterns \cite{Goldman:2004,Bougie:2005}.
The Swift-Hohenberg   theory is based on the equation for the
order parameter $\psi$,
\begin{equation}
\partial_t\psi
=[\epsilon-(\nabla^2+k_0^2)^2]\psi-\psi^3+\eta({\bf x},t),
\label{SHE}
\end{equation}
where $\epsilon$ is the bifurcation parameter, $k_0$ is the
wavenumber corresponding to the most unstable perturbations, and
$\eta$ is the Gaussian $\delta$ correlated noise term with
intensity $F$. The Swift-Hohenberg theory predicts that noise
offsets the bifurcation value of the control parameter from the
mean-field value $\epsilon_{MF}=0$ to the critical value
$\epsilon_c\propto F^{2/3}$. Furthermore, the Swift-Hohenberg theory describes
the transition to the linear regime which is expected to work far
away from the bifurcation point for small noise intensity when the
magnitude of noise-excited modes scales as
$|\epsilon-\epsilon_c|^{-1/2}$, while the time coherence of
fluctuations and the amplitude of spectral peaks decays as
$|\epsilon-\epsilon_c|^{-1}$. Fitting the Swift-Hohenberg equation
(\ref{SHE}) to match the transition in vibrated granular layer,
\textcite{Goldman:2004,Bougie:2005} found a good agreement with molecular dynamics
simulations  and experiments (see, e.g.,
Fig.\ref{fig_swinney_noise}).  Interestingly, the magnitude of the
fitted noise term in Eq.(\ref{SHE}) $F\approx 3.5\cdot 10^{-3}$ turned
out to be an order of magnitude greater than for convective
instability in a fluid near a critical point \cite{Oh:2003}. This
discrepancy could stem from the fact that the Swift-Hohenberg theory, developed
for ordinary fluids,  is formally valid for the second-order phase
transition, whereas in granular system the transition to square
patterns is of the first order type.
Consequently, the nonlinear terms can be important near the
transition point and may distort the scaling for the noise
amplitude.

\begin{figure}
\caption{ Comparison between the Swift-Hohenberg theory and
experiment for noise peak intensity (a), total noise power (b) and
the correlation time (c). Symbols - experiment, solid lines -
Swift-Hohenberg theory, dashed lines - linear theory for small
noise magnitude, from \textcite{Goldman:2004}.}
\label{fig_swinney_noise}
\end{figure}

There have been attempts to connect patterns in vibrated layers
with the phenomenon of granular ``thermoconvection''. Since
high-frequency vibration in many aspects is similar to ``hot''
wall, it was argued that one should expect granular temperature
gradients,  density inversion,  and, consequently convection
instability similar to that observed in heated from below liquid
layers. The  theoretical analysis based on granular hydrodynamic
equations (\ref{mass}),(\ref{momentum}),(\ref{energy}) supports
the existence of a convective instability in a certain range of
parameters \cite{He:2002,Khain:2003}. Multiple convection roles
were observed in molecular dynamics simulations
\cite{Sunthar:2001,Paolotti:2004}. However, the experiments are
not conclusive enough \cite{Wildman:2001}. In particular it
appears very hard to discriminate between convection induced by
vibration and convective flows induced by walls, see e.g.
\cite{Pak:1993,Garcimartin:2002}.

Vibrated bottom plate is not the only way to induce parametric
patterns in thin granular layers. \textcite{Li:2003} demonstrated
that periodically modulated airflow through a shallow fluidized
bed also produces interesting patterns in the granular layer which
oscillate at half the driving frequency (Fig.
\ref{fig_fluidbed_patterns}). While the physical mechanism of
interaction between the airflow and grains is quite different from
the collisional energy transfer in vibrated containers,
phenomenological models based on the principal symmetry of the
problem should be able to describe the gas-driven granular layer
as well. In case of the parametric Ginzburg-Landau model Eq.
(\ref{eq0a}), the order parameter would correspond to the amplitude
of the subharmonic component of the surface deformation, and the
driving term would be related to the amplitude of the flow
modulation. Moreover, variations of the mean flow rate act similar
to the variations of the gravitational acceleration in the
mechanical system, which may give an additional means to control
the state of the system.

\begin{figure}
\caption{Select patterns in a shallow fluidized bed with
periodically modulated air flow for different flow parameters,
from \textcite{Li:2003}} \label{fig_fluidbed_patterns}
\end{figure}

\section{Patterns in gravity-driven dense granular flows}
\label{sec:dense}

In this Section we overview theoretical models for various
pattern-forming instabilities in dense gravity-driven granular
flows.

\subsection{Avalanches in thin granular layers}

Gravity-driven particulate flows are a common occurrence in nature
(dune migration, erosion/deposition processes, land slides,
underwater gravity currents and coastal geomorphology) and in
various industrial applications having to do with handling
granular materials, including their storage, transport, and
processing. One of the most spectacular (and often very dangerous)
forms of gravity-driven granular flows is the avalanche.
Avalanches occur spontaneously when the slope of the granular
material exceeds a certain angle (static angle of repose) or they
can be initiated at somewhat smaller angle by applying a finite
perturbation. Laboratory  studies of avalanches are often carried
out in rotating drums (see below) or in a chute geometry when a
layer of sand is titled at a certain  fixed angle.
\textcite{Daerr:1999} conducted experiments with a thin layer of
granular matter on sticky (velvet) inclined plane, see Fig.
\ref{fig_daerr2}. Surprising diversity of avalanche behavior was
observed in this seemingly simple system: triangular avalanches
developed in thin layers ($h$ is the layer thickness) and for
small inclination angles $\phi$, whereas in thicker layers or
steeper angles $\phi$ the avalanches assumed balloon shaped with
upper edge of the avalanche propagating up-hill, see Fig.
\ref{fig_daerr1}. According to
\textcite{Rajchenbach:2002a,Rajchenbach:2003} the rear front of
the balloon-like avalanche propagates uphill with the velocity
roughly one half of the downhill velocity of the head front, and
the velocity of the head is also two times larger than the
depth-averaged flow velocity. The stability diagram is outlined in
Fig. \ref{fig_daerr2}: a granular layer is stable below solid line
(so-called $h_{stop}$ limit according to
\textcite{Pouliquen:1999}), spontaneous avalanching was observed
above the dashed line. Between dashed and solid lines the layer
exhibits bistable behavior: finite perturbation can trigger an
avalanche, otherwise the layer remains stable. The dotted line
with $\times$-symbols indicates the transition between triangular
and balloon avalanches.

\begin{figure}
\caption{Stability diagram for avalanches in thin granular layers,
se also Fig. \ref{fig_daerr1},  from \textcite{Daerr:1999}}
\label{fig_daerr2}
\end{figure}

\subsubsection{Partially fluidized flows}
\label{subsubsec:part}

The avalanche dynamics described above is an example
of a wide class of {\em partially fluidized} granular flows. In such flows
part of grains flows past each other while other grains maintain static contacts with
their neighbors. The
description of such flows still represents a major challenge for the
theory.  In particular, one is faced with the problem of constructing
the constitutive relation for the stress tensor $\bsigma$.
In dense quasi-static flows a significant part of the stresses
is transmitted through quasi-static contacts between particles as
compared with short collisions in dilute flows.

Stimulated by the non-trivial avalanche dynamics in experiments by
\textcite{Daerr:1999,Daerr:2001}, \textcite{Aranson:2001,Aranson:2002b} suggested
a generic continuum description of partially fluidized granular flows.
 According to this theory, the ratio of the static part $\bsigma^s$ to the fluid part
$\bsigma^f$ of the full stress tensor is controlled by the order
parameter $\rho$. The order parameter is scaled in such a way that
in granular solid $\rho = 1$ and in well developed flow (granular
liquid) $\rho \to 0$. On the ``microscopic level" the order
parameter is defined as a fraction of the number of static (or
persistent) contacts of the particles $Z_s$ to total number of the
contacts $Z$, $\rho= \langle Z_s/Z \rangle$ within a mesoscopic
volume which is large with respect to the particle size but small
compared with characteristic size of the flow.

Due to  a strong dissipation in dense granular flows the order
parameter $\rho$  is assumed to obey purely relaxational dynamics
controlled by the Ginzburg-Landau-type equation for the generic
first order phase transition,
\begin{equation}
\frac{D\rho}{Dt}=D\nabla^2\rho-\frac{\partial {F}(\rho, \delta)}{\partial
\rho}. \label{GL2}
\end{equation}
Here $D$ is the diffusion coefficient.    ${F}(\rho, \delta)$ is a free
energy density which is postulated to have two local minima
at $\rho=1$ (solid phase) and $\rho=0$ (fluid phase) to
account for the bistability near the solid-fluid transition.

The relative stability of the two phases is controlled by the
 parameter $\delta$ which in turn is determined
by the stress tensor.  The simplest assumption consistent with the
Mohr-Coloumb yield criterion is to take it as a function of
$\phi=\max |\sigma_{mn}/\sigma_{nn}|$, where the maximum is sought
over all possible orthogonal directions $m$ and $n$ (we consider
here only two-dimensional formulation of the model, an objective
three dimensional  generalization was recently proposed by
\textcite{Gao:2005}). Furthermore, there are two angles which
characterize the fluidization transition in the bulk of granular
material, an internal friction angle $\tan^{-1}\phi_1$ such that
if $\phi > \phi_1$ the static equilibrium is unstable, and the
``dynamic repose angle'' $\tan^{-1}\phi_0$ such that at
$\phi<\phi_0$, the ``dynamic'' phase $\rho=0$, is unstable. Values
of $\phi_0$ and $\phi_1$ depend on microscopic properties of the
granular material, and in general they do not coincide.
\textcite{Aranson:2001,Aranson:2002b} adopted the simple
algebraic form of the control parameter $\delta$,
\begin{equation}
\delta=(\phi^2-\phi_0^2)/(\phi_1^2-\phi_0^2). \label{delta00}
\end{equation}

Order parameter equation (\ref{GL2}) has to be augmented by
boundary conditions. While this is a complicated issue in general, a
simple but meaningful choice is to take no-flux boundary conditions
at free surfaces and smooth walls, and solid phase condition $\rho=1$
near sticky or rough walls.

\begin{figure}
\caption{Comparison of theoretical and experimental phase
diagrams. Lines obtained from theory, symbols depicts experimental
data from Ref. \protect \cite{Daerr:1999}. Solid line and circles
limit the range of existence of avalanches, line and triangles
correspond to the linear stability boundary of the static chute,
and the  line and crosses denote the boundary between triangular
and balloon avalanches. Inset: Schematic representation of a chute
flow geometry, from \textcite{Aranson:2001,Aranson:2002b}}
\label{fig_ar4}
\end{figure}

For the flow of thin granular layers on inclined planes  Eqs.
(\ref{momentum}), (\ref{GL2}) can be simplified. Using the no-slip boundary
condition at the bottom and no-flux condition at the top of the layer
and fixing the lowest-mode
structure of the order parameter in the direction perpendicular to
the bottom of the chute ($z=0$, see Inset to Fig. \ref{fig_ar4}),
$\rho=1-A(x,y) \sin(\pi z/h)$, $h$ is the local layer thickness,
$A(x,y)$ is slowly-varying function, one obtains equations
governing the evolution of thin layer
\cite{Aranson:2001,Aranson:2002b}:
\begin{eqnarray}
\partial_t h &=&  - \alpha \partial_x(h^3 A)
+ \frac{\alpha}{\phi}  \nabla \left(h^3 A \nabla h \right) ,
\label{conser} \\
\partial_t A  &=& \lambda A+ \nabla^2_\perp A
+\frac{8(2-\delta) }{3 \pi} A^2 -\frac{3 }{4}  A^3 \label{A1}
\end{eqnarray}
where $\nabla^2_\perp = \partial_x^2+\partial_y^2$,
$\lambda=\delta-1-\pi^2/4 h^2$,  $\alpha \approx 0.12 \mu^{-1} g
\sin \bar \varphi$, $\mu$ is the shear viscosity, $\bar \varphi$
is the chute inclination, $\phi=\tan \bar \varphi$. Control
parameter $\delta$ includes correction due to change in the local
slope $\delta=\delta_0+\beta h_x$, $\beta \approx
1/(\phi_1-\phi_0)\approx 1.5-3$ depending on the value of $\phi$.
The last term in Eq. (\ref{conser}) is also due to change of local
slope angle $\varphi$ and is obtained from the expansion $\varphi
\approx \bar \varphi + h_x$. This term is responsible for the
saturation of the avalanche front slope (without it the front
would be arbitrarily steep). While it was not included in original
publications \cite{Aranson:2001,Aranson:2002b}, this term is
important for large wavenumber cut-off of long-wave instability
observed by \textcite{Forterre:2003}, see  Sec. \ref{subsec:inst}.
Numerical and analytic solutions of Eqs. (\ref{conser}),(\ref{A1})
exhibit strong resemblance with experiment: triangular avalanches
in thin layers and balloon-like avalanches in thicker layers, see
Fig. \ref{fig_ar3}. The corresponding phase diagram agrees
quantitatively  with an experimental one having only one fitting
parameter (viscosity $\mu$), Fig. \ref{fig_ar4}.

\begin{figure}
\caption{ Sequence of  images demonstrating the evolution of a
triangular avalanche (a-c) and up-hill avalanche (d-f) obtained
form numerical solution of  Eqs. (\protect
\ref{A1}),(\ref{conser}), from
\textcite{Aranson:2001,Aranson:2002b}} \label{fig_ar3}
\end{figure}

In subsequent work \cite{Aranson:2002b}, this theory was
generalized to other dense shear granular flows including flows in
rotating drums, two- and three-dimensional shear cells, etc.  The
model also was tested in soft-particle molecular dynamics
simulations \cite{Volfson:2003a,Volfson:2003b,Volfson:2004}.
\textcite{Orpe:2005} used the partial fluidization model of
\cite{Aranson:2001,Aranson:2002b,Volfson:2003a} for the
description of velocity profiles three-dimensional shear flows in
a rotating drum. The comparison between experimental data and theory
shows that the partial fluidization model describes reasonably
well entire velocity profile and the flow rheology, however
experimental methods for independently estimating the order
parameter model are needed. \textcite{Gao:2005} recently developed
an objective (coordinate system independent) formulation of the
partial fluidization theory which allows for the straightforward
generalization to three-dimensional systems.

\subsubsection{Two-phase  flow approach  of granular avalanches}
\label{subsubsec:two}

Another approach treating near-surface granular flows as
two-phase systems  was developed by a number of authors, see e.g.
\cite{Bouchaud:1994,Bouchaud:1995,Boutreux:1998,Douady:1999,Mehta:1994,Khakhar:2001}
and many others. For review on recent models of surface flows see
\cite{Aradian:2002}. All these models distinguish rolling
and static phases of granular flow described by the set of coupled
equations for the evolution of thicknesses of both phases, $R$ and $h$, respectively. The phenomenological
theory by \textcite{Mehta:1994,Bouchaud:1994,Bouchaud:1995} (often
called BCRE theory) provides an intuitive description of the flow.
In shallow granular layers, even simpler
depth-averaged granular hydrodynamic equations
(so-called Saint-Venant models) often provides quite accurate
description, see
\cite{Savage:1989,Douady:1999,Khakhar:2001,Lajeunesse:2004}.

In the most general and compact form the BCRE theory can be
represented by a pair of equations for evolution of $R$ and $h$,
\begin{eqnarray}
\partial_t h &=&  \Gamma(h,R) \label{bcre1} \\
\partial_t R &=& v_d \partial_x R
 - \Gamma(h,R) \label{bcre2}
\end{eqnarray}
where $ \Gamma(h,R)$ is the {\it exchange term}, or a conversion
rate between rolling and static grains, and $v_d$ is the downhill
grain velocity. Physical meaning of the BCRE model is very simple:
Eq. (\ref{bcre1}) expresses the increase in the height due to
deposition of rolling grains, and Eq. (\ref{bcre2}) describes
advection of rolling fraction by the flow with velocity $v_d$ and
depletion due to conversion to static fraction. The limitations and
generalizations of the BCRE model are discussed by
\textcite{Boutreux:1998,Aradian:2002}.

\textcite{Douady:2002} applied the following two-phase model to
describe avalanches in thin granular layers:
\begin{eqnarray}
 \partial_t h +2 U\partial_x h
&=&  \frac{g}{\bar \Gamma} \left(\tan \phi - \mu(h) \right)  \label{douad1} \\
\partial_t \zeta  +2 U\partial_x h
&=& 0 \label{douad2}
\end{eqnarray}
where $\zeta=R+h$ is the position of free surface, $U$ is
depth-averaged velocity of the flow. In addition to BCRE model
Eqs. (\ref{douad1}),(\ref{douad2}) include two phenomenological
functions:  $\bar \Gamma$ characterizes the mean velocity
gradient of a single bead on incline, and $\mu(h)$ describes
depth-dependent friction with the bottom. According to
\textcite{Douady:2002} a three-dimensional version of Eqs.
(\ref{douad1}),(\ref{douad2}) describes transition from triangular
to uphill avalanches, however details of the transition
depend sensitively on the choice of functions $\bar \Gamma$ and
$\mu(h)$.

Depth-averaged description in the form of Eqs.
(\ref{douad1}),(\ref{douad2}) was used by
\textcite{Borzsonyi:2005a} to address the difference between
shapes of avalanches for sand and glass particles in a chute flow.
The authors reduced Eqs. (\ref{douad1}),(\ref{douad2}) to the
modified Burgers equation
\begin{equation}
\partial_t h + a(h) \partial_x h = \mu(h)  \partial_x^2 h \label{burg2}
\end{equation}
where function $a(h) \sim h^{3/2} $  and effective viscosity $\mu
\sim \sqrt h$. This description connects avalanches with the
``Burgers'' shocks. Eq. (\ref{burg2}) implies that all avalanches
will eventually decay, in contrast to experiments indicating that
only small avalanches decay whereas large avalanches grow and/or
form stationary waves \cite{Daerr:1999,Daerr:2001a}.  This
discrepancy is likely due to the fact that reduction of the full
model (\ref{douad1}),(\ref{douad2}) to the single equation
(\ref{burg2}) does not take into account the bistable nature of
granular flows.

While two-phase description of granular flow is simple and rather intuitive,
it can be problematic when a clear-cut separation between rolling
and static phases is absent, especially near the onset of
motion. The order parameter approach can be more appropriate in this
situation. Furthermore, the two-phase equations  can be derived from
the  partial fluidization model  described in the Sec.
\ref{subsubsec:part}  as a sharp-interface limit of the continuum order
parameter model \cite{Aranson:2002b}.

\subsubsection{Avalanche shape}
On the basis of simple kinematic considerations \textcite{Rajchenbach:2002b}
suggested an analytic expression for the shape of triangular and
balloon-like avalanches. For the balloon-like avalanches the shape is given
by the envelope of the expanding circles with the center
drifting downhill:
\begin{equation}
x^2+\left(y-2 \bar v t +\frac{5}{2} \bar v \tau\right)^2 = \left(
\frac{1}{2} \bar v \tau \right)^2, \;\; 0 < \tau< t \label{raj1}
\end{equation}
where $\bar v$ is the velocity of the rear front. For the
triangular avalanches the shape is given by the envelope of
dilating ellipses
\begin{equation}
\left( \frac{ \bar v x}{2 v_\perp} \right)^2+\left(y-
\frac{3}{2}\bar v t \right)^2 = \left( \frac{1}{2} \bar v t
\right)^2.  \label{raj2}
\end{equation}
Here $v_\perp $ is perpendicular velocity. While these
heuristic relations are consistent with experimental
observation, see Fig. \ref{fig_raj1}, their connection to
continuum dynamical models of granular flows remains to be understood.

\begin{figure}
\caption{Top: Total area overrun by the avalanche (solid line),
compared with experimental image from \cite{Daerr:1999}. Bottom:
superimposition of avalanche boundaries given by Eq. (\protect
\ref{raj2}) for three different moments of time, from
\textcite{Rajchenbach:2002b}} \label{fig_raj1}
\end{figure}

\subsection{Statistics of avalanches and sandpile model}
\label{sec:SOC}
It is well known that in real sandpiles avalanches can vary widely in size.
The wide distribution of scales in real avalanches stimulated \textcite{Bak:1987} to
introduce a ``sandpile cellular automaton'' as
a paradigm model for the {\em self-organized criticality}, the
phenomenon which occurs in slowly driven non-equilibrium spatially extended systems
when they asymptotically reach a critical state characterized by a
power-law distribution of event sizes. The set of rules which constitute
the sandpile model is very simple. Unit size ``grains" are dropped one by one on a
one-dimensional lattice in random places and form vertical stacks.
If a local slope (the difference  between heights of two neighboring
stacks) exceeds a certain threshold value, a grain hops from the higher
to the lower stack. This may trigger an ``avalanche" of subsequent hops
until the sandpile returns to the stable
state.
After that another grain is dropped and the relaxation process
repeats.
The size of an avalanche
is determined by the number of grains set into motion by adding a single
grain to a sandpile.  This model asymptotically reaches a critical state in which the
mean angle is equal to the critical slope, and avalanches have a universal power-law
distribution of sizes, $P(s)\propto s^{-\alpha}$ with $\alpha\approx 1.5$.

The relevance of this model and its generalizations to the
real avalanches is still the matter of debate.  The sandpile model by \textcite{Bak:1987} is
defined via a single repose angle and so its asymptotic behavior has the
properties of the critical state for a second-order phase transition.
Real sandpiles are characterized by two angles of repose and thus
exhibit features of the first-order phase transition.
Moreover, concept of self-organized criticality is related to a power-law distribution of avalanche sizes, thus
reliable experimental verification of self-organized criticality requires
accumulation of very large statistics of avalanche events  and a large-scale experimental setup.
Finite size effects should strongly affect the power-law behavior.

Experiments with avalanches in slowly rotating drums
\cite{Jaeger:1989,Rachenbach:2000} and chute flows
\cite{Lemieux:2000} do not confirm the scale-invariant
distribution of avalanches.  In other experiments with large
mono-disperse glass beads dropped on a conical sandpile
\textcite{Costello:2003} claimed existence of  the self-organized
criticality with $\alpha\approx 1.5$. Characteristics of the size
distribution depended on the geometry of the sandpile, physical
and geometrical properties of grains, and the way the grains are
dropped on the pile, contrary to the universal concept of
self-organized critical behavior. Self-organized criticality was
also claimed in the avalanche statistics in three-dimensional pile
of anisotropic grains (long rice), however a smaller scaling
exponent $\alpha\approx 1.2$ was measured for the avalanche size
distribution \cite{Aegerter:2004}. Interestingly, rice piles were
observed to demonstrate roughening dynamics of their surface as
the distribution of active sites in the self-organized critical
state shows a self-affine structure with the fractal exponent
$d_B=1.85$ \cite{Aegerter:2004}. This is consistent with the
theoretically predicted mapping between self-organized criticality
and roughening observed for example in Kardar-Parizi-Zhang model
\cite{Paczuski:1996}.

One can argue  that
real sandpiles should  not exhibit self-organized criticality in a strict
 sense due to hysteresis and the existence of two
critical  repose angles. However, since the difference between the angles is relatively small, one cannot
exclude power-law type behavior in the {\it finite range}  of avalanche sizes.
 This circumstance  possibly explains significant scatter in
 experimental results and scaling exponents for avalanche size distribution and
  the dependence on grain shape and material properties.

\subsection{Instabilities in granular chute flows}
\label{subsec:inst}

Granular chute flows exhibit a variety of pattern-forming instabilities,
including fingering \cite{Pouliquen:1997,Malloggi:2005}, longitudinal
vortices \cite{Forterre:2001,Forterre:2002,Borzsonyi:2005}, long surface
waves \cite{Forterre:2003}, segregation and stratification
\cite{gray97c,makse97}, etc.

\textcite{Pouliquen:1997} studied experimentally a granular chute
flow on a rough inclined plane. Experiments performed with
polydisperse sand particles demonstrated fingering instability of
the  front propagating down the slope, similar to that observed in
fluid films flowing down inclined plane
\cite{Zhou:2005,Troian:1989}.  However, similar experiments with
smooth monodisperse glass beads exhibited no instability. The
authors argued that the instability was due to a flow-induced  size
segregation in a polydisperse granular matter. The segregation
indeed was found near the avalanche front. However, similar
experiments \cite{Malloggi:2005,Malloggi:2005b} showed a fingering
front instability without a significant size segregation. Thus,
the question of the mechanism of fingering instability is still
open.

Experiments by
\textcite{Forterre:2001,Forterre:2002,Borzsonyi:2005} show the
development of longitudinal vortices in rapid chute flows, see
Fig. \ref{fig_forterre}. The vortices develop for large
inclination angles and large flow rates in the regime of
accelerating flow when the flow thickness decreases and the mean
flow velocity increases along the chute. \textcite{Forterre:2001}
proposed an explanation of this phenomenon in terms of  granular
``thermoconvection''. Namely, rapid granular flow has a high shear
near the rough bottom which leads to the local increase of
granular temperature and consequently creates a density inversion.
In turn, the density inversion trigger a convection instability
similar to that in ordinary fluids. The critical instability
wavelength $\lambda_C$ is determined by the depth of the layer $h$
(in experiment $\lambda_c \approx 3 h$).

In a subsequent work \textcite{Forterre:2002} studied the formation of
longitudinal vortices and the stability of granular chute flows  in the
framework of granular hydrodynamics
Eqs.(\ref{mass})-(\ref{energy}). The inverse density profile
appears when a heuristic  boundary condition at the bottom relating
slip velocity and heat flux is introduced. Steady-state
solution of Eqs.(\ref{mass})-(\ref{energy}) indeed yields an inverse
density profile (Fig. \ref{fig_forterre2}) which turns out to be
unstable with respect to short-wavelength perturbations for large
flow velocities, see Fig. \ref{fig_forterre3}. While the linear stability
analysis captured many important features of the phenomenon, there
are still open questions.
The stability analysis was performed for the steady flow whereas the instability
occurs in the regime of accelerating flow. Possibly due to this assumption
 the linear stability analysis yielded
oscillatory instability near the onset of vortices, whereas for
the most part, the vortices appear to be steady.
Another factor which is ignored in the theory is the air drag. The
high flow velocity in the experiment (about 1-2 m/sec) is of the
order of the terminal velocity of an individual grain in air, and
therefore air drag may affect the granular flow.

\begin{figure}
\caption{Density profiles $\nu(z)$ as function of distance form
the chute bottom $z$ for different values of mean flow velocity,
from \textcite{Forterre:2002}} \label{fig_forterre2}
\end{figure}

\begin{figure}
\caption{Phase diagram in mean density ($\bar \nu$) and flow
thickness ($h$) plane  delineating different flow instabilities.
Smaller $\bar nu$ corresponds to faster flow, from
\textcite{Forterre:2002}} \label{fig_forterre3}
\end{figure}

\begin{figure}
\caption{Long-surface wave instability observed in flow of sand
down rough incline, from \textcite{Forterre:2003}}
\label{fig_forterre4}
\end{figure}

\begin{figure}
\caption{Experimental dispersion relation for the long surface wave
instability. Shown spatial growth rate as function of the
frequency of forcing wave, from \textcite{Forterre:2003}}
\label{fig_forterre5}
\end{figure}

\textcite{Forterre:2003}  presented an experimental study of the
long-surface-wave instability developing  in  granular  flows on
a rough inclined plane, Fig. \ref{fig_forterre4}. This instability
was known from previous studies \cite{Savage:1979,Davies:1990},
however no precise characterization of the instability had been
performed. \textcite{Forterre:2003} measured the threshold and the
dispersion relation of the instability by imposing a controlled
perturbation at the entrance of the flow and measuring its
evolution down the slope, see Fig. \ref{fig_forterre5}. The results
are compared with the prediction of a linear stability analysis
conducted in the framework of depth-averaged Saint-Venant-type
equations similar to those described in Sec. \ref{subsubsec:two}:
\begin{eqnarray}
\label{fort1}
 \partial_t h &+& \partial_x ( u h) = 0 \\
\partial_t( u h)&+& \alpha \partial_x(u^2 h)
= \left(\tan \theta - \mu(u,h)- \partial_x h
\ \right)   g h \cos(\theta) \nonumber
\end{eqnarray}
where $h$ is local thickness, $\theta$ is inclination angle, $u$
is depth-averaged flow velocity, $\mu(h,u)$ is a function
describing effective depth and velocity dependent bottom friction,
$\alpha \sim O(1)$ is a
constant determined by the velocity profile within the layer. According to
\textcite{Forterre:2003}, the instability is similar to the long-wave
instability observed in classical fluids but with characteristics
that can dramatically differ due to the specificity of the granular
rheology.  The theory is able to predict quantitatively the stability
threshold and the phase velocity of the waves but fails to describe the
observed cutoff of the instability at high wavenumbers.  Most likely, one
needs to include higher order terms, such as $\partial_x^2 h$ in the
first Eq. (\ref{fort1}) in order to account for the cutoff.

\begin{figure}
\caption{The growth rate of small perturbations $\sigma$ vs
wavelength $k$ derived from Eqs. (\ref{conser}),(\ref{A1}) for
$\beta=2$, $\alpha=0.025, \delta=1.1$. Instability occurs near
$h_{stop}$ curve in Figs. \ref{fig_daerr2},\ref{fig_ar4} ($h=2.9$)
and disappears with further increase of $h$.} \label{fig_sigma}
\end{figure}

\begin{figure}
\caption{Typical profiles of hight $h$ and order parameter $A$ in
the regime of long-surface wave instability for $\beta=2$,
$\alpha=0.025, \delta=1.1$. Starting from generic initial
conditions $h=h_0, A=const$ plus small noise, a sequence of
avalanches develops.} \label{fig_longwave}
\end{figure}

The order parameter theory based on Eqs.
(\ref{conser}),(\ref{A1}) also reproduces the long-surface
wave instability. Furthermore, linearizing Eqs. (\ref{conser}),(\ref{A1})
near the steady flowing solution $A=A_0+ \tilde a \exp[ \sigma t
+ ikx], h = h_0 + \tilde h \exp[ \sigma t + ikx]$, after simple
algebra one obtains that the growth rate of linear perturbations
$\sigma$ is positive only in a band restricted by some critical
wavenumber and only in the vicinity of $h_{stop}$, see Fig.
\ref{fig_sigma}. With the increase of $h$, i.e. the granular flux,
the instability disappears, in agreement with experiments.
The nonlinear saturation of the instability results in the
development of a sequence of avalanches, which is generally
non-periodic, see Fig. \ref{fig_longwave}. The structure  shows
slow coarsening due to merging of the avalanches. This
instability is a possible candidate mechanism of the formation of
inhomogeneous deposit structure behind the front of an
avalanche.

\textcite{Conway:2003} studied free-surface waves in granular chute
flows near a frictional boundary. The experiments
showed that the sub-boundary circulation driven by the velocity
gradient plays an important role in the pattern formation, suggesting
a similarity between wave generation in granular and fluid flows.

A Kelvin-Helmholtz-like shear instability in chute flows was
observed by \textcite{Goldfarb:2002},  when two  streams of sand
flowing on an inclined plane with different velocities were in
side-by-side contact with each other. For sufficiently high chute
angles and shear rates the interface remains flat. The instability
of the interface develops when the chute angle and/or the shear
rate is reduced. This instability has been reproduced in
soft-particle molecular dynamics simulations by
\textcite{Ciamarra:2005} who also observed that in a polydisperse
medium this instability leads to grain segregation (See below
Sect. \ref{sec:segr}).

\subsection{Pattern-forming instabilities in rotating cylinders}
\label{sect_drum} Granular media in rotating horizontal cylinders
(drums) often show behavior similar to chute flows. For very small
rotation rates (as defined by small Froude number $Fr=\omega^2
R/g$, where $\omega$ is the angular velocity of drum rotation and
$R$ its radius), well separated in time avalanches occur when the
slope of the free surface exceeds a certain critical angle
$\theta_c$ whereby diminishing this angle to a smaller {\em static
repose angle} $\theta_s$
\cite{Rachenbach:1990,Jaeger:1989,Tegzes:2002,Tegzes:2003}. The
difference between $\theta_c$ and $\theta_s$ is usually a few
degrees. At an intermediate rotation speed, a continuous flow of
sand emerges instead of discrete avalanches through a hysteretic
transition, similar to the transition in chute flows  at large
rates of grain deposition \cite{Lemieux:2000}. In the bulk, the
granular material rotates almost as a solid body with some
internal slipping.  As moving grains reach the free surface they
slide down within a thin near-surface layer \cite{Zik:1994} (see
sketch in Fig. \ref{fig_drum}). The surface has a nearly flat
shape; the arctangent of its average slope defines the
so-called {\em dynamic angle of repose} $\theta_d$.

There are various models
addressing the nature of the transition from discrete avalanches
to the continuum flow. \textcite{Linz:1995} proposed a phenomenological
model based on a system of equation for the angle of repose $\phi$
and mean flow velocity $v$
\begin{eqnarray}
\dot v & =&  g \left ( \sin \phi - k(v)  \cos \phi)
\right)  \chi (\phi,v) \nonumber \\
\dot \phi & = & \bar \omega - a v \label{linz}
\end{eqnarray}
where $\bar \omega$ is the rotation frequency of the drum,
$k(v)=b_0 + b_2 v^2$ is the velocity dependent friction coefficient,
 $\chi
(\phi,v)$ is some cut-off function, and $a,b_0,b_2$ are parameters
of the model. Despite the simplicity, the model yields
qualitatively correct transition from discrete avalanches to
continuous flow with the increase of rotation rate $\bar \omega$,
and also  predicts logarithmic relaxation of the free surface
angle in the presence of vibration.

The transition from avalanches to flow  naturally arises in the
framework of the partial fluidization theory, \cite{Aranson:2002b}. In
this case one can derive a system of coupled equations for the parameter
$\delta$ (which is related to  the surface local angle $\phi$, see Eq.
(\ref{delta00})) and the width of fluidized layer $z_0$,
\begin{eqnarray}
\partial_t z_0 & =&  \partial_s^2 z_0 +F(z_0,\delta)-\bar v \partial_s z_0  \nonumber \\
\partial_t \delta & = & \bar \omega + \partial_s^2 J \label{drum1}
\end{eqnarray}
where $s$ is the coordinate along the slope of the granular surface
inside the drum,
$J=f(z_0)$ is the downhill flux of grains, $\bar v$ is averaged
velocity in flowing layer, and functions $F, f$ and $v_0$ are
derived from  Eq. (\ref{GL2}). This model bears resemblance to the
BCRE-type models of surface granular flows which were applied to
rotating drums by \textcite{Khakhar:1997,Makse:1999}.

Eqs. (\ref{drum1}) exhibit stick-slip type oscillations of the
surface angle for slow rotation rates and a hysteretic transition to
a steady flow for larger rates. Eqs. (\ref{drum1}) yields the
following scaling for the width of the flowing layer $z_0$ in the
middle of the drum vs rotation frequency: $z_0 \sim \bar
\omega^{2/3} $, which is consistent with experiment
\cite{Tegzes:2002,Tegzes:2003}. After integration over $s$
Eqs. (\ref{drum1}) can be reduced to a system of two coupled
equations for averaged drum angle $\langle \delta \rangle$  and
averaged flow thickness $\langle z_0 \rangle $ somewhat similar to
the model of \textcite{Linz:1995}.

Granular flows in long rotating drums under certain conditions also
exhibit fingering instability \cite{Shen:2002,Fried:1998}.  Similarity
between fingering in rotating drums and chute flows
\cite{Forterre:2002} suggests that mechanisms described in the Section
\ref{subsec:inst} can be responsible for this effect, see also Section
\ref{sec:segr}.

\section{Models of granular segregation}
\label{sec:segr}

One of the most fascinating features of heterogeneous (i.e.,
consisting of different distinct components) granular materials is
their tendency to segregate under external agitation rather
than to mix, as one would expect from the naive entropy
consideration.  This property is ubiquitous in Nature (see, e.g.
\cite{iverson97}) and has important technological implications
\cite{cooke76}. In fact, some aspects of segregation of small
and large particles can be understood on equilibrium
thermodynamics grounds \cite{asakura58}. Since the excluded volume
for small particles around large ones becomes smaller when large
grains clump together, separated state possesses lower entropy.
However, granular systems are driven and strongly dissipative, and
this simple equilibrium argument can only be applied
qualitatively. The granular segregation is more widespread than it
would be dictated by thermodynamics. In fact, any variation in
mechanical properties of particles (size, shape, density,
surface roughness, etc.) may lead to their segregation. At least for
bi-disperse rapid dilute flows the granular segregation can be
rigorously treated in the framework of kinetic theory of
dissipative gases, see Subsec. \ref{subsec:hydrodynamic}.
\textcite{Jenkins:2002a} employed kinetic theory for a binary
mixture for spheres or disks in gravity and derived a simple
segregation criterion based on the difference of partial pressures
for each type of particles  due to the difference in size and/or
mass.

\begin{figure}[ptb]
\caption{Granular stratification in a flow down heap, from
\textcite{makse97}.} \label{fig_stratified_heap}
\end{figure}

Segregation has been observed in most flows of granular mixtures,
including granular convection \cite{knight93}, hopper flows
\cite{makse97,gray97c,Samadani:1999,Samadani:2001},  flows in
rotating drums \cite{Zik:1994,hill97,Choo:1997}, and even in
freely cooling binary granular gases \cite{Catuto:2004}.
Segregation among large and small particles due to shaking has
been termed ``Brazil nut effect'' \cite{rosato87}. The phenomenon
of granular segregation was discovered long time ago, and several
``microscopic'' mechanisms have been proposed to explain its
nature, including inter-particle collisions \cite{brown39},
percolation \cite{williams76}, and others.  In certain cases,
separation of grains produces interesting patterns. For example,
if a binary mixture of particles which differ both in size {\em
and} in shape is poured down on a plane, a heap which consists of
thin alternating layers of separated particles is formed
\cite{gray97c,makse97}, see Fig. \ref{fig_stratified_heap}.
Rotating of mixtures of grains with different sizes in long drums
produces well separated bands of pure mono-disperse particles
\cite{Zik:1994,hill97,Choo:1997,chicarro97}, Fig.
\ref{fig_axial_bands}.  In this Section we only address models of
pattern formation due to segregation (stratification and banding),
without discussing other manifestations of granular segregation.

\subsection{Granular stratification}
Granular stratification occurs when a binary mixture of particles
with different physical properties is slowly poured on a plate
\cite{gray97c,makse97,koeppe98}. More specifically, it occurs when larger
grains have additionally larger roughness resulting in a larger
repose angle and the flux of falling particles is small enough to
cause intermittent avalanches down the slopes of the heap. The
basic mechanism of stratification is related to the avalanches
acting as {\em kinetic sieves} \cite{savage88,savage93}. During an
avalanche, voids are continuously being created
within flowing near-surface layer, and small particles are more likely to
fall into them. This creates a downward flux of smaller particles
which is compensated by the upward flux of larger particles in
order to maintain a zero total particle flux across the flowing
layer. Other models of granular segregation in a thin flowing
layer \cite{Khakhar:1997,dolgunin98,Khakhar:1999} lead to a
similar result. Each avalanche leads to the formation of a new pair of
layers in which the grains of different sorts are separated (see
Fig. \ref{fig_stratified_heap}). This pair of layers grows from
the bottom of the pile by upward propagation of a kink at which
small particles are stopped underneath large ones. However, when
the larger particles were smooth and small particles were rough,
instead of stratification only large scale segregation with small
particles near the top and large particles near the bottom was
observed.


\begin{figure}[ptb]
\caption{Cellular automata model of granular stratification, from
\textcite{makse97b}.} \label{fig_makse1}
\end{figure}

\textcite{makse97,makse97b} proposed a cellular automata model
which generalized the classical sandpile model \cite{Bak:1987} (see
Section \ref{sec:SOC}). In
this model, a sandpile is built on a lattice, and rectangular
grain have identical horizontal size but different heights (see Fig.
\ref{fig_makse1}a). Grains are released at the top of the heap
sequentially, and they are allowed to roll down the slope. A
particle would become rolling if the local slope (defined as the
height difference between neighboring columns) exceeds the repose
angle. To account for difference in grain properties, four
different repose angles $\theta_{\alpha\beta}$
were introduced for grains of type $\alpha$ rolling on a
substrate of type $\beta$ ($\alpha,\beta\in \{1,2\}$ where 1 and 2
stand for small and
large grains, respectively). Normally, $\theta_{21}<\theta_{12}$
because of the geometry (small grains tend to get trapped by large
grains), and one-component repose angles usually lie within this
range, $\theta_{21}<\theta_{11},\theta_{22}<\theta_{12}$. However
the ratio of $\theta_{11},\theta_{22}$ depends on the relative
roughness of the grains. For
$\theta_{21}<\theta_{11}<\theta_{22}<\theta_{12}$ (large grains
are more rough), the model yields stratification in agreement with
experiment (Fig. \ref{fig_makse1}b). If, on the other hand,
$\theta_{22}<\theta_{11}$ (which corresponds to smaller grains
being more rough), the model yields only large-scale segregation:
large particles collect at the bottom of the sandpile.

This physical model can also be recast in the form of continuum
equations \cite{boutreux96,makse97b} which generalize the
single-species BCRE model of surface granular flows
\cite{Bouchaud:1994} (see Section \ref{sec:dense}):
\begin{eqnarray}
\partial_t R_\alpha&=&-v_\alpha\partial_x
R_\alpha+\Gamma_\alpha, \label{makse1}
\\
\partial_t h&=&-\sum_\alpha\Gamma_\alpha,
\label{makse2}
\end{eqnarray}
where $R_\alpha(x,t), v_\alpha$ are the thickness and velocity of
rolling grains of type $\alpha$, $h(x,t)$ is the instantaneous
profile of the sandpile, and $\Gamma_\alpha$ characterizes
interaction between the rolling grains and the substrate of static
grains. In the same
spirit as in the discrete model, the interaction function
$\Gamma_\alpha$ is chosen in the form
\begin{equation}
\Gamma_\alpha=\left\{\begin{array}{l}\gamma_\alpha
[\theta_{l}-\theta_\alpha(\phi_\beta)]R_\alpha
\\\gamma_\alpha\phi_\alpha[\theta_{l}-\theta_\alpha(\phi_\beta)]R_\alpha
\end{array}.\right.
\label{makse3}
\end{equation}
Here $\phi_\alpha(x,t)$ is the volume fraction of grains of type
$\alpha$, and $\theta_{l}=-\partial_x h$ is the local
slope of the sandpile. This form of the interaction terms implies
that the grains of type $\alpha$ become rolling if the local slope
exceeds the repose angle $\theta_\alpha(\phi_\beta)$ for this type
on a surface with composition $\phi_\beta(x,t)$. Assuming that the
generalized repose angles $\theta_\alpha(\phi_{\beta})$ are linear
functions of the concentration
\begin{eqnarray}
\theta_1(\phi_2)=(\theta_{12}-\theta_{11})\phi_2+\theta_{11},
\\
\theta_2(\phi_2)=(\theta_{12}-\theta_{11})\phi_2+\theta_{21}.
\end{eqnarray}
Eqs. (\ref{makse1})-(\ref{makse3}) possess a stationary solution in
which the heap is separated into two regions where
$\theta_2(\phi_2)<\theta<\theta_1(\phi_2)$ and
$\theta<\theta_2(\phi_2)<\theta_1(\phi_2)$. This solution
corresponds to small grains localized near the top and small
grains near the bottom with a continuous transition between the
two regions. However, \textcite{makse97} showed that this
stationary solution  is unstable if
$\delta=\theta_{22}-\theta_{11}>0$ and gives rise to the
stratification pattern.

Similar effect of stratification patterns was observed
experimentally in a thin slowly rotating drum which is more than
half filled with a similar binary mixture \cite{gray97c}, see Fig.
\ref{fig_gray97}. Periodic avalanches, occurring in the drum, lead
to formation of strata by the same mechanism described above.

\begin{figure}[ptb]
\caption{Granular avalanche-induced stratification in rotating
drum observed for low rotation rates, from \textcite{gray97c}.}
\label{fig_gray97}
\end{figure}

\subsection{Axial segregation in rotating drums}
\label{sec_axial} The most common system in which granular
segregation is studied is a rotating drum, or a partially filled
cylinder rotating around its horizontal axis (see Section
\ref{sect_drum}). When a polydisperse mixture of grains is rotated
in a drum, strong {\em radial segregation} usually occurs within
just a few revolutions. Small and rough particles aggregate to the
center ({\em core}) of the drum, large and smooth particles rotate
around the core (see Figs. \ref{fig_radial} and \ref{fig_drum}).
Since there is almost no shear flow in the bulk, the segregation
predominantly occurs within a thin fluidized near-surface
layer. For long narrow drums with the length much exceeding the
radius, radial segregation is often followed by the {\em axial
segregation} occurring at later stages (after several hundred
revolutions) when the angle of repose of small particles exceeds
that of large particles. As a result of axial segregation, a
pattern of well segregated bands is formed \cite{Zik:1994,hill97}
(see, e.g., Fig. \ref{fig_axial_bands}) which slowly merge and
coarsen. Depending on the rotation speed, coarsening can either
saturate at a certain finite bandwidth at low rotation speeds when
discrete avalanches provide granular transport \cite{Frette:1997}
or at higher rotation rates in a continuous flow regime it can
lead to a final state in which all sand is separated in two bands
\cite{Zik:1994,Fiodor:2003}.

The axial segregation has been well known in the engineering
community, it was apparently first observed by Oyama in 1939
\cite{oyama39}. The mechanism of axial segregation is apparently related
to the different friction properties of grains which lead to different
{\em dynamical angles of repose}. The latter are defined as the angle of
the slope in the drum corresponding to continuous flow regime, however in
real drums the free surface has a more complicated S-shape
\cite{Zik:1994,Elperin:1998,Makse:1999,Orpe:2001}. According to
\textcite{Zik:1994} (see also \cite{Levine:1999}), if there is a local
increase in concentration of particles with higher dynamic repose angle,
the local slope there will be higher, and that will lead to a local bump near
the top of the free surface and a dip near the bottom. As the particles
tend to slide along the steepest descent path, more particles with
higher repose angle will accumulate in this location, and the
instability will develop.
\textcite{Zik:1994} proposed a
quantitative continuum model of axial segregation based on the equation for the
conservation equation for the relative concentration of the two
components (``glass'' and ``sand''),
$c(z,t)=(\rho_A-\rho_B)/(\rho_A+\rho_B)$,
\begin{equation}
\partial_t c=-{C\over
\rho_T}(\tan\theta_A-\tan\theta_B)\partial_z
(1-c^2)\langle(1+y_x^2){y_z\over y_x}\rangle. \label{zik}
\end{equation}
Here $x$ and $z$ are Cartesian horizontal coordinates across and
along the axis of the drum, $y(x,z,t)$ describes the
instantaneous free surface inside the drum,
$\rho_T=\rho_A+\rho_B$, $C$ is a constant related to gravity and
effective viscosity of granular material in the flowing layer. The
term in angular brackets denotes the axial flux of the glass beads
averaged over the cross-section of the drum. The profile of the free
surface in turn should depend  on $c(z,t)$. If
$\langle(1+y_x^2)y_c/y_x\rangle<0$, linearization of
Eq.(\ref{zik}) leads to the diffusion equation with negative
diffusion coefficient which exhibits segregation instability with
growth rate proportional to the square of the wavenumber.  It is
easy to see that the term in angular brackets vanishes for a
straight profile $y_x=const(x)$. However, for the experimentally
observed S-shaped profile of the free surface \textcite{Zik:1994}
calculated that the instability condition is satisfied when the
drum is more than half full. While experiments show that axial
segregation in fact observed even for less than 50\% filling
ratio, the model gives a good intuitive picture for the mechanism
of the instability.

Recent experiments \cite{hill97,Choo:1997,Choo:1998,Fiodor:2003}
have revealed interesting new features of axial segregation.
\textcite{hill97} performed magnetic resonance imaging  studies
\cite{hill97} which demonstrated that in fact the bands of larger
particles usually have a core of smaller particles. More recent
experiments by \textcite{Fiodor:2003} showed that small particles
formed a shish kebab-like structure with bands connected by a
rod-like core, while large particles formed disconnected rings.
\textcite{Choo:1997,Choo:1998} found that at early stages, the
small-scale perturbations propagate across the drum in both
directions (this was clearly evidenced by the experiments on the
dynamics of pre-segregated mixtures \cite{Choo:1997}), while at
later times more long-scale static perturbations take over and
lead to the emergence of quasi-stationary bands of separated
grains (see Fig. \ref{fig_morris}). The slow coarsening process
can be accelerated in a drum of a helical shape \cite{Zik:1994}.
Alternatively, the bands can be locked in an axisymmetrical  drum
with the radius modulated along the axis \cite{Zik:1994}.

\begin{figure}[ptb]
\caption{Space-time diagram of the surface of long rotating drum
demonstrating oscillatory size segregation. The plot shows the
full length of the drum and extends over 2,400 sec, or 1,850
revolutions. Black bands correspond to 45-250 $\mu$m   black sand
and white bands correspond to 300-850 $\mu$m table salt,  from
\textcite{Choo:1997}.} \label{fig_morris}
\end{figure}

In order to account for the oscillatory behavior of axial
segregation at the initial stage,
\textcite{Aranson:1999b,Aranson:1999c} generalized the model of
\textcite{Zik:1994}. The key assumption was that besides the
concentration difference, there is an additional slow variable
which is involved in the dynamics.
\textcite{Aranson:1999b,Aranson:1999c} conjectured that this
variable is the instantaneous slope of the granular material
(dynamic angle of repose) which unlike Eq.(\ref{zik}) is not slaved
to the relative concentration $c$, but obeys its own dynamics. The
equations of the model read
\begin{eqnarray}
\partial_t c=-\partial_{z}(-D\partial_{z}c + g(c)\partial_{z}\theta),
\label{conc}
\end{eqnarray}
\begin{eqnarray}
\partial_t\theta&=&\alpha(\Omega - \theta + f(c)) +
D_\theta\partial_{zz}\theta + \gamma\partial_{zz} c.
\label{theta}
\end{eqnarray}
The first term in the r.h.s. of Eq.(\ref{conc}) describes diffusion flux (mixing),
and the second term describes differential flux of particles due
to the gradient of the dynamic repose angle.  This term is
equivalent to the r.h.s. of Eq.(\ref{zik}) with a particular
function $g(c)=G_0(1-c^2)$. For simplicity, the constant $G_0$ can
be eliminated by rescaling of distance $x \to x/ \sqrt{G_0}$.
The sign $+$  before this term means that the particles with the
larger  static repose angle are driven towards greater dynamic
repose angle. This differential flux gives rise to the segregation
instability. Since this segregation flux vanishes with $g(c)$
$|c|\to 1$ (which correspond to pure $A$ or $B$ states), it
provides a natural saturation mechanism for the segregation
instability.

Parameter $\Omega$ in the second equation is the normalized angular velocity
of the drum rotation, and $f(c)$ is the static angle of repose
which is an increasing function of the relative
concentration \cite{koeppe98} (for simplicity it can be assumed linear,
$f(c)= F + f_0 c  $).  The constant $F$ can be eliminated by the
substitution  $\theta \to \theta -F$. The first term in the r.h.s. of
Eq.(\ref{theta}) describes the local dynamics of the repose angle
($\Omega$ increases the angle, and $-\theta+f(c)$ describes the
equilibrating effect of the surface flow), and the term $
D_\theta\partial_{xx}\theta$ describes axial diffusive relaxation.
The last term, $ \gamma \partial_{xx}c$, represents the
lowest-order non-local contribution from an inhomogeneous
distribution of $c$ (the first derivative $\partial_x c$ cannot be
present  due to reflection symmetry $x\to -x$). This term gives
rise to the transient oscillatory dynamics of the binary mixture.

Linear stability analysis of a homogeneous state $c=c_0;
\theta_0=\Omega+f_0 c_0$ reveals that for $g_0f_0 >\alpha D$
long-wave perturbations are unstable, and if
$g_0\gamma>(D_\theta-D)^2/4$, short-wave perturbations oscillate
and decay (two eigenvalues $\lambda_{1,2}$ are complex conjugate
with negative real part), see Fig. \ref{fig_disp_curve}.  This
agrees with the general phenomenology observed by
\textcite{Choo:1997} both qualitatively and even quantitatively
(Fig. \ref{fig_disp_curve}b).  The results of direct numerical
solution of the full model (\ref{conc}),(\ref{theta}) are
illustrated by Fig. \ref{fig_axial_theor}. It shows that
short-wave initial perturbations decay and give rise to more
long-wave non-oscillatory modulation of concentration which
eventually leads to well-separated bands. At long times (Fig.
\ref{fig_axial_theor}b) bands exhibit slow coarsening with the
number of bands decreasing logarithmically with time (see also
\textcite{Frette:1997,levitan98,Fiodor:2003}). This scaling
follows from the exponentially weak interaction between interfaces
separating different bands \cite{Aranson:1999c,fraerman97}.

\begin{figure}[ptb]
\caption{Dispersion relation $\lambda(k)$ for segregation
instability (left) and comparison of the frequency of band
oscillations $Im(\lambda)$ with experiment (right), from
\textcite{Aranson:1999c}.} \label{fig_disp_curve}
\end{figure}

\begin{figure}[ptb]
\caption{Space-time diagrams demonstrating initial band
oscillations and consequent coarsening, from
\textcite{Aranson:1999c}.} \label{fig_axial_theor}
\end{figure}

While these continuum models of axial segregation showed a good
qualitative agreement with the data, recent experimental
observations demonstrate that the theoretical understanding of
axial segregation is far from complete \cite{Ottino:2000}. The
interpretation of the second slow variable as the local dynamic
angle of repose implies that in the unstable mode the slope and
concentration modulation should be in phase, whereas in the
decaying oscillatory mode, these two fields have to be shifted in
phase.  Further experiments \cite{khan04} showed that while the
in-phase relationship in the asymptotic regime holds true, the
quadrature phase shift in the transient oscillatory regime is not
observed. That lead \textcite{khan04} to hypothesize that some
other slow variable other than the angle of repose (possibly
related to the core dynamics) may be involved in the transient
dynamics. However, so far experiments failed to identify which
second dynamical field is necessary for oscillatory transient
dynamics, so it remains an open problem. Another recent
experimental observation by \textcite{Khan:2005} suggested that instead
of normal diffusion assumed in Eqs.(\ref{conc}),(\ref{theta}), a
slower subdiffusion of particles in the core takes place, $\langle
r\rangle\sim t^\gamma$ with the
scaling exponent $\gamma$ close to 0.3. The most plausible explanation is
that the apparent subdiffusive behavior is in fact a manifestation
of {\it nonlinear} concentration diffusion which can be described by equation
\begin{equation}
\partial_t c = \partial_z D(c) \partial_z c.
\label{conc_n}
\end{equation}
For example, for the generic concentration-dependent diffusion
coefficient  $D \sim c$, the asymptotic scaling behavior of the
concentration $c(z,t) $ is given by the self-similar function $c
\sim F(z/t^\alpha)/t^\alpha$ for $t \to \infty$ with the scaling
exponent $\alpha = 1/3$ close to 0.3 observed experimentally.
Experimentally observed scaling function $F(x/t^\alpha)$ appears
to be consistent with that of Eq. (\ref{conc_n}) except for the
tails of the distribution where $c \to 0$ and the assumption $D
\sim c$ is possibly violated. Normal diffusion behavior
corresponding to $D=const$ and $\alpha=1/2$ is in strong
disagreement with the experiment.


\textcite{Newey:2004} conducted studies of axial segregation in
ternary mixtures of granular materials. It was found that for
certain conditions bands of ternary mixtures oscillate axially. In
contrast to the experiments of \textcite{Choo:1997,Choo:1998}, the
oscillations of bands appear spontaneously from initially mixed
state, which  strongly indicates the supercritical {\it oscillatory
instability}. While in binary mixtures the oscillations have the
form of periodic mixing/demixing of bands, in the ternary mixtures
the oscillations are in the form of periodic band displacements.
It is likely that the  mechanism of band oscillations in ternary
mixtures is very different from that of binary mixtures. One of
possible explanations could be that the third mixture component
provides an additional degree of freedom necessary for
oscillations. To demonstrate that we write phenomenological
equations for the concentration differences $C_A=c_1-c_2$ and
$C_B=c_2-c_3$, where $c_{1,2,3}$ are the individual
concentrations. By analogy with Eq. (\ref{conc}) we write the
system of coupled equations for the concentration differences
$C_{A,B}$ linearized near the fully mixed state:
\begin{eqnarray}
\partial_t C_A & =& D_A \partial_z^2 C_A + \mu_A \partial_z^2 C_B,
\nonumber \\
\partial_t C_B & =& D_B \partial_z^2 C_B + \mu_B \partial_z^2 C_A.
\label{concab}
\end{eqnarray}
If the cross-diffusion terms have opposite signs, i.e. $\mu_A
\mu_B < 0$, the concentrations $C_{A,B}$ will exhibit oscillations
in time and in space. Obviously this mechanism is intrinsic to
ternary systems and has no counterpart in binary mixtures.

Parallel to the theoretical studies, molecular dynamics
simulations have been performed
\cite{shoichi98,rapaport02,Taberlet:2004}. Simulations allowed researchers
to probe the role of material parameters which would be difficult to
access in laboratory experiments. In particular,
\textcite{rapaport02} addressed the role of particle-particle and
wall-particle friction coefficients separately.  It was found that
the main role is played by the friction coefficients between the
particles and the cylinder walls: if the friction coefficient
between large particles and the wall is greater than that for
smaller particles, the axial segregation always occur irrespective
of the ratio of particle-particle friction coefficients. However,
if the particle-wall coefficients are equal, the segregation may
still occur if the friction among large particles is greater than
among small particles. \textcite{Taberlet:2004}  studied axial
segregation in a system of grains made of identical material
differing only by size. The simulations
revealed rapid oscillatory motion of bands, which is  not
necessarily related to the slow band appearence/disappearence observed
in experiments of \textcite{Choo:1997,Choo:1998,Fiodor:2003}.

A different type of discrete element modelling of axial
segregation was proposed by \textcite{Yanagita:1999}. This model
builds upon the lattice-based sandpile model and replaces a rotating
drum by a three-dimensional square lattice. Drum rotation is
modelled by correlated displacement of particles on the lattice:
particles in the back are shifted upward by one position, and the
particles at the bottom are shifted to fill the voids. This
displacement steepens the slope of the free surface, and once it
reaches a critical value, particles slide down according to the
rules similar to the sandpile model of \textcite{Bak:1987} but taking
into account different critical slopes for different particles.
This model despite its simplicity reproduced both radial
and axial segregation patterns and therefore elucidated
the critical components needed for adequate description of the
phenomenon.

\subsection{Other examples of granular segregation}
\label{subsubsec:other} As we have seen in the previous Section,
granular segregation occurs in near-surface shear granular flows,
such as in silos, hoppers,  and rotating drums. However, other
types of shear granular flows may also lead to segregation.  For
example, Taylor-Couette flow of granular mixtures between two
rotating cylinders leads to formation of Taylor vortices and then
in turn to segregation patterns \cite{shinbrot04a}, see Fig.
\ref{fig_shinbrot04}.

\begin{figure}[ptb]
\caption{Granular Taylor vortices observed in vertically
rotating air-fluidized cylinder filled with binary mixture. Left
image depicts entire cylinder height and width, and right image
shows the dependence of  concentration of small particles along
the bed height, from \textcite{shinbrot04a}.}
\label{fig_shinbrot04}
\end{figure}

\textcite{Pouliquen:1997} observed granular segregation in a thin
granular flow on an inclined plane. In this case, segregation
apparently occurs as a result of an instability in which
concentration mode is coupled with hydrodynamic mode. As a result,
segregation occurs simultaneously with a fingering instability of
the avalanche front (Fig. \ref{fig_pouliquen97b}). As an implicit
evidence of this relation between segregation and fingering
instability, \textcite{Pouliquen:1997} found that mono-disperse
granular material does not exhibit fingering instability. However,
other experiments \cite{Shen:2002} indicate that in other
conditions (more rapid flows), fingering instability may occur
even in flows of mono-disperse granular materials. Thus, the
segregation is likely a consequence rather than the primary cause
of the fingering instability.

An interesting recent example of pattern formation caused by granular
segregation in a horizontally shaken layer of binary granular
mixture was presented by
\textcite{Mullin:2000,Mullin:2002,Reis:2002}. After several
minutes of horizontal shaking with frequency 12.5 Hz and
displacement amplitude 1 mm (which corresponds to the acceleration
amplitude normalized by gravity $\Gamma=0.66$), stripes were
formed orthogonal to the direction of shaking. The width of the
stripes was growing continuously with time as $d\propto t^{0.25}$,
thus indicating slow coarsening (Fig. \ref{fig_mullin}).  This
power law is consistent with the diffusion-mediated mechanism of
stripe merging. \textcite{Reis:2002} argued on the basis of
experimental results on patterned segregation in horizontally
shaken layers that the segregation bears features of the
second-order phase transition. Critical slow-down was observed
near the onset of segregation. The order parameter is associated
with the combined filling fraction $C$, or the layer compacity,
\begin{equation}
C=\frac{N_s A_s +N_l A_l}{S}
\end{equation}
where $N_{s,l} $ are numbers of particles in each species, $A_{s,l}
$ are projected two-dimensional areas of the respective individual
particles, and $S$ is the tray area. \textcite{Ehrhard:2005}
proposed a simple numerical model to describe this phenomenon of
segregation in horizontally vibrated layers. The model is based on
a two-dimensional system of hard disks of mass $m_\alpha$ and radius
$R_\alpha$ ($\alpha=1,2$ denote the species)
\begin{equation}
m_\alpha \dot {\bf v}_{\alpha i} = -\gamma_i \left({\bf v}_{\alpha
i}-{\bf v}_{tray}(t) \right) + \zeta_{\alpha i} (t)
\end{equation}
where ${\bf v}_i$ is the particles velocity ${\bf v}_{tray}(t) =
A_0 \sin (\omega t) $ is oscillating tray velocity, $\gamma$
provides linear damping, and $\zeta_{\alpha i}$ is Gaussian white
noise acting independently on each disk. The model reproduced
segregation instability and subsequent coarsening of stripes. More
realistic discrete element simulations were recently performed by
\textcite{Ciamarra:2005}. In these simulations a binary mixture of
round disks of identical sizes but two different frictions with the
bottom plate (in fact, velocity-dependent viscous drag was
assumed), separated in alternating bands perpendicular to the
oscillation direction irrespectively on initial conditions: both
random mixed state and separated along the direction of
oscillations state were used. Using particles of the same size
eliminated the thermodynamic ``excluded volume'' mechanism for
segregation, and the authors argued that the mechanism at work is
related to the dynamical shear instability similar to the
Kelvin-Helmholtz instability in ordinary fluids. It was confirmed
by a numerical observation of the interfacial instability  when
two monolayers of grains with different friction constant were
placed in contact along a flat interface parallel to the direction
of horizontal oscillations. Similar instability is apparently
responsible for ripple formation \cite{Scherer:1999,Stegner:1999}.

\textcite{Pooley:2004} proposed theoretical description of this
experiment based on continuum model for periodically-driven two
isothermal ideal gases which interact through frictional force. It
was shown analytically that segregated stripes form spontaneously
above critical forcing amplitude. While the model reproduces the
segregation instability, apparently it does not exhibit coarsening of
stripes observed in the experiment. Moreover, applicability of the
isothermal ideal gas model to this experiment where the particles
are almost at rest is an open question.

Similar coarsening effect in granular segregation in a
particularly simple geometry was studied by \textcite{aumaitre01}.
They investigated the dynamics of a monolayer of grains of two
different sizes in  a dish shaken in a horizontal ``swirling'' motion.
They observed that large particles tend to aggregate near the
center of the cavity surrounded by small particles. The
qualitative explanation of this effect follows from simple
thermodynamic considerations (see above). Indeed, direct tracing
of particle motion showed that the pressure in the area near the
large particles is smaller than outside. But small particles do
not follow the gradient of pressure and assemble near the center
of the cavity because this gradient is counterbalanced by the
force from large particles.  The inverse of force acting on large
particles leads to their aggregation near the center of the
cavity. \textcite{aumaitre01} proposed a more quantitative model
of segregation based on the kinetic gas theory and found
satisfactory agreement with experimental data.

\textcite{Burtally:2002} studied spontaneous separation of
vertically vibrated mixtures  of particles of similar sizes but
different densities (bronze and glass spheres). At low frequencies
and at sufficient vibrational amplitudes, a sharp boundary between
the lower layer of glass beads and the upper layer of the heavier bronze
spheres was observed. At higher frequencies, the bronze particles emerge as a
middle layer separating upper and lower glass bead layers.  The
authors argue that the effect of air on the granular motion is
a relevant  mechanism of particle separation. A somewhat similar
conclusion was achieved by \textcite{Moebius:2001} in experiments
with vertically-vibrated column of grains containing a large
``intruder'' particle.


\textcite{Fiodor:2003,Arndt:2005} performed detailed experiments
on axial segregation in slurries, or bi-disperse grain-water
mixtures. A mixture of two types of spherical glass beads of two sizes were
placed in a water-filled tube at the volume ratio 1:2.
Authors found that both rotation rate and filling fraction
play an important role in band formation. Namely, bands  are less
likely to form at lower fill levels (20-30\%) and slower rotation
rates (5-10 rpm). They mostly appear near the ends of the drum. At
higher fill levels and rotation rates, bands form faster, and
there are more of them throughout the drum.
\textcite{Fiodor:2003,Arndt:2005} also studied the
relation between the bands visible on the surface, and the core of
small beads, and found that for certain fill levels and rotation
speeds, the core remains prominent at all times, while in other
cases the core disappears completely between bands of small
particles. They also observed an interesting oscillatory
instability of interfaces between bands at high rotation speeds.
All these phenomena still await theoretical modelling.

\section{Granular materials with complex interactions}
\label{sec:comp}

\subsection{Patterns in solid-fluid mixtures}

Presence of interstitial fluid significantly complicates the
dynamics of granular materials.  Hydrodynamic flows lead to
the viscous drag and anisotropic long-range interaction between
particles.  Even small amounts of
liquid leads to cohesion among the particles which can have a
profound effect on macroscopic properties of granular assemblies
such as angle of repose, avalanching, ability to segregate, etc.
(see for example Sec. \ref{sec_axial} and
\textcite{Samadani:2000,Samadani:2001,Tegzes:2002,Li:2005}).

In this Section we will discuss the case when the volume fraction
of fluid in the two-phase system is large, and the grains are
completely immersed in fluid. This is relevant for many industrial
applications, as well as for geophysical problems such as
sedimentation, erosion, dune migration, etc.

One of the most technologically important examples of particle-laden flows is
a fluidized bed. Fluidized beds  have been widely used
since German engineer Fritz Winkler invented the first
fluidized bed for coal gasification in 1921. Typically, a vertical
column containing granular matter is energized by a flow of gas or
liquid. Fluidization occurs when the drag force exerted by the
fluid on the granulate exceeds gravity. A uniform fluidization,
the most desirable regime for most industrial applications, turns
out to be prone to bubbling instability: bubbles of clear fluid
are created spontaneously at the bottom, traverse the granular layer and destroy the
uniform state \cite{Jackson:2000}. Instabilities in fluidized beds
is an active area of research in the engineering community,
see \cite{Jackson:2000,Gidaspow:1994,Kunii:1991}. A shallow fluidized
bed shows many similarities with mechanically vibrated layers, see
Section \ref{sec:multil}. In particular, modulations of airflow
studied by \textcite{Li:2003} result in formation of subharmonic
square and stripe patterns (see Fig. \ref{fig_fluidbed_patterns})
similar to those in mechanically-vibrated systems
\cite{Melo:1994,Melo:1995,Umbanhowar:1996}.

Wind and water driven granular flows play important roles in
geophysical processes. Wind-blown sand forms dunes and beaches.
The first systematic study of airborne (or aeolian) sand transport
was conducted by R. Bagnold during Wold War II, see
\textcite{Bagnold:1954} who identified two primary mechanisms of
sand transport: saltation and creep,  and proposed the first
empiric relation for the sand flux $q$ driven by wind shear stress
$\tau$:
\begin{equation}
q= C_B \frac{\nu_{a}} {g}  \sqrt{\frac{d}{D}}  u_*^3
\label{Bagnold}
\end{equation}
where $C_B=const$, $\nu_{a} $ is air density, $d$ is the grain
diameter, $D=0.25$ mm is a reference grain size, and $u_*= \sqrt{
\tau/\nu_a}$ is wind friction velocity. Later many refinements of
Eq.(\ref{Bagnold}) were proposed, see e.g. \cite{Pye:1991}.

\textcite{Nishimori:1993} proposed a simple theory which describes formation of  ripples as well as  dunes.
The theory  is based on a lattice model which incorporates separately
saltation and creep processes. The model operates with the height of sand at each lattice side at
discrete time $n$, $h_n(x,y)$. The full time step includes two substeps. The saltation substep is described as
\begin{eqnarray}
\bar h_n (x,y) &=& h_n(x,y) -q                \\
\bar h_n(x+L(h(x,y)),y)&=& h_n(x+L(h_n(x,y)),y)+q \nonumber
\label{nishim1}
\end{eqnarray}
where $q$ is the height of grains being transferred from one (coarse grained) position $(x,y)$ to the other
position $(x+L,y)$ on the lee side (wind is assumed blowing in the positive $x$ direction),
$L$ is the flight length in one saltation which characterizes the wind strength.
It is assumed that $q$ is conserved. Since the saltation length $L$ depends on multiple factors,
the following simple approximation is accepted
\begin{equation}
L=L_0 + b h_n(x,y)
\label{nishim2}
\end{equation}
with $L_0$ measuring wind velocity and  $b=const$. The creep substep involves  spatial averaging
over neighboring sites in order to describe the surface relaxation due to gravity,
\begin{eqnarray}
&&h_{n+1}(x,y)=\bar h_n(x,y)+ \\
&&D\left[
\frac{1}{6} \sum_{NN}\bar h(x,y) + \frac{1}{12} \sum_{NNN}\bar h(x,y)-\bar h(x,y)
\right],   \nonumber
\label{nishim3}
\end{eqnarray}
where $ \sum_{NN}$ and $ \sum_{NNN}$ denote summation over the
nearest neighbors and next nearest neighbors correspondingly,  and
$D=const$ is the surface relaxation rate. Despite its simplicity,
simulation of the model reproduced formation of ripples and
consequently arrays of barchan (crescent shaped) dunes, see Fig.
\ref{fig_nishimori}. \textcite{Nishimori:1993} found that above
certain threshold an almost linear relation holds between the
selected wavelength of the dune pattern and the ``wind strength"
$L$.

\begin{figure}
\caption{Sand ripple pattern (top panel) and barchan dunes (lower
panel) obtained from simulations of Eqs.
(\ref{nishim1})-(\ref{nishim3}) \textcite{Nishimori:1993}}
\label{fig_nishimori}
\end{figure}

In the long-wave limit Eqs. (\ref{nishim1})-(\ref{nishim3}) can be reduced  to more traditional
continuum models considered below.

In the continuum description of the evolution of the sand surface, the  profile $h$ is governed by
the mass conservation equation
\begin{equation}
\nu_{s} \partial_t h = - \nabla {\bf q} \label{mass1},
\end{equation}
where $\nu_s$ is the density of sand and ${\bf q}$ is the sand
flux. In order to close Eqs. (\ref{Bagnold}),(\ref{mass1}),
several authors proposed different phenomenological relations
between shear stress at the bed surface ${\bf \tau}$ and the
height $h$, see e.g.
\cite{Nishimori:1993,Kroy:2002a,Kroy:2002b,Hersen:2004,Prigozhin:1999,Andreotti:2002}.

There are many theories generalizing \textcite{Nishimori:1993}
approach, see e.g. \cite{Caps:2001}. \textcite{Prigozhin:1999}
described the evolution of dunes by a system of two equations
similar to the BCRE model discussed earlier in Sec.
\ref{sec:dense} \cite{Bouchaud:1994,Bouchaud:1995}. One equation
describes the evolution of the local height
 $h$ while
another equation describes the density $R$ of particles rolling above the stationary
sand bed profile (reptating particles),
\begin{eqnarray}
 \partial_t h &=&  \Gamma(h,R)-f \\
\partial_t R &=& -\nabla {\bf J} +{\bf Q} -
\Gamma(h,R) \label{prig}
\end{eqnarray}
where $\Gamma$ is the rolling-to-steady sand transition rate,
${\bf J}$ is the horizontal projection of the flux of rolling
particles, ${\bf Q}$ accounts for the influx of falling reptating
grains, and $f$ is the erosion rate.  With an appropriate
choice of rate functions $\Gamma, f, Q$ and $J$, Eqs. (\ref{prig})
can reproduce many observed features of dune formation, such as
initial instability of flat state, asymmetry of the dune profiles,
coarsening and interaction of dunes, etc., see Fig. \ref{fig_prigozhin}.

\begin{figure}
\caption{Interaction and coarsening of one-dimensional dune
system, from \textcite{Prigozhin:1999}} \label{fig_prigozhin}
\end{figure}

\begin{figure}
\caption{Evolution of two barchan dunes described by Eqs.
(\ref{Bagnold}), (\ref{mass1}). Small dune is undersupplied and
eventually shrinks, from \textcite{Hersen:2004}}
\label{fig_hersen1}
\end{figure}

Thus, simplified models such as
\cite{Nishimori:1993,Prigozhin:1999,Kroy:2002a} have been successful in explaining
many features of individual dune growth and evolution, see Fig.
\ref{fig_hersen1}. However we should note that up to date none of the dune models
have been able to
address satisfactorily the wavelength selection in large-scale dune fields
\cite{Hersen:2004}.

The phenomenon qualitatively similar to the dune formation occurs
in an oscillatory fluid flow above a granular layer: sufficiently
strong flow oscillations produce so-called vortex ripples on the
surface of the underlying granular layer. These ripples are
familiar to any beachgoer.  Vortex ripple formation was first
studied by \textcite{Ayrton:1910,Bagnold:1946}, and recently by
\textcite{Stegner:1999,Scherer:1999} and others. It was found that
ripples emerge via a hysteretic transition, and are
characterized by a near-triangular shape with slope angles close to
the repose angle. The characteristic size of the ripples $\lambda$
is directly proportional to the displacement amplitude of the
fluid flow $a$ (with a proportionality constant $\approx 1.3$) and
is roughly independent on the frequency.

\textcite{Andersen:2001}
introduced order parameter models for describing the dynamics of
sand ripple patterns under oscillatory flow based on the
phenomenological  mass transport law  between adjacent ripples.
The models predict the existence of a stable band of wave numbers
limited by secondary instabilities and coarsening of small
ripples, in agreement with experimental observations.

\textcite{Langlois:2005} studied underwater ripple formation on a
two-dimensional sand bed sheared by viscous fluid. The sand
transport is described by generalization of Eq. (\ref{Bagnold})
taking into account both the local bed shear stress  and the local
bed slope. Linear stability analysis revealed that ripple
formation is attributed to a growing longitudinal mode. The weakly
nonlinear analysis taking into account resonance interaction of
only three unstable modes revealed a variety of steady
two-dimensional ripple patterns drifting along the flow at some
speed.

Experiments in dune formation have been recently
performed in water \cite{Betat:1999}. While water-driven and wind driven
dunes and ripples have similar shape, the underlying physical processes
are likely not the same due to a different balance between gravity and
viscous drag in air and water.

\begin{figure}
\caption{Fingering instability of planar avalanche in the
underwater flow, figure on the left zooms on individual finger,
from  \textcite{Malloggi:2005}} \label{fig_malloggi}
\end{figure}

Spectacular erosion patterns in sediment granular layers were
observed in experiments with underwater flows
\cite{Daerr:2003,Malloggi:2005}. In particular, a fingering
instability of flat avalanche fronts was observed, see Fig.
\ref{fig_malloggi}. These patterns are remarkably similar to those
in thin films on inclined surfaces, both with clear and
particle-laden  fluids \cite{Troian:1989,Zhou:2005}. In the
framework of lubrication approximation the evolution of fluid film
thickness $h$ is described by the following dimensionless equation
following from the mass conservation law:
\begin{equation}
\partial_t h + \nabla \cdot \left\{  \left[ h^3
\nabla \nabla^2 h \right] - \bar D h^3 \nabla h \right \} +
\partial_x h^3 = 0 \label{film}
\end{equation}
where dimensionless parameter $\bar D$ is inversely proportional
to water surface tension. The instability occurs for small $\bar
D$ values, i.e. in the large surface tension limit. However,
despite visual similarity the physical mechanism leading to this
fingering instability is not obvious: in fluid films the
instability is driven (and stabilized) by the surface tension,
whereas in the underwater granular flow fluid surface tension
plays no role.

\textcite{Duong:2004} studied formation of periodic arrays of
knolls in a slowly rotating horizontal cylinder filled with granular
suspension, see Fig. \ref{fig_doung}. The solidified sediment
knolls co-exist with freely circulating fluid. The authors applied
variable viscosity fluid which formally allows simultaneous
treatment of solid and liquid phase. In this model the effective
flow viscosity $\mu_s$ diverges at the solid packing fraction
$\phi_{rcp}$,
\begin{equation}
\mu_s= \frac{\mu_0}{(1-\phi/\phi_{rcp})^b}
\end{equation}
where $\mu_0$ is the clear fluid viscosity and $b$ is an empirical
coefficient. The model qualitatively reproduces the experiment, see
Fig. \ref{fig_doung}. An interesting question in this context is
whether there is a connection to the experiment by
\textcite{Shen:2002} where somewhat similar structures were
obtained for the flow of ``dry'' particles in a horizontally
rotating cylinder.

\begin{figure}
\caption{Upper panel: Self-supporting knolls formed in water/glass
beads suspension in a horizontally rotating cylinder (side and end
views). Lower panel: Computational results: schematics of the flow
(a); the height of computed knoll structures (b,c), from
\textcite{Duong:2004}} \label{fig_doung}
\end{figure}

As it was mentioned in Sec. \ref{subsubsec:other},
\textcite{Conway:2004} reported  that an air-fluidized vertical
column of bi-disperse granular media sheared between
counter-rotating cylinders exhibits formation of nontrivial vortex
structure strongly reminiscent of Taylor vortices in conventional
fluid, see e.g. \cite{Andereck:1986}. Authors argue that vortices
in fluidized granular media, unlike Taylor vortices in fluid,  are
accompanied by the novel segregation-mixing mechanism specific for
granular systems, see Fig. \ref{fig_shinbrot04}. Interestingly, no
vortices were observed in a similar experiment in Couette geometry
with monodisperse glass beads \cite{Losert:2000}.

\textcite{Ivanova:1996} studied patterns in a
horizontal cylinder filled with sand/liquid mixture and subject to
horizontal vibration. For certain vibration parameters standing
wave patterns were observed at the sand/liquid interface. Authors
argue that these wave patterns are similar to the Faraday ripples found at
liquid/liquid interface under vertical vibration.

\subsection{Vortices in vibrated rods}

In Section \ref{sec:multil} we reviewed instabilities and collective
motion in mechanically vibrated layers. In most
experiments the particle shape was not important. However,
strong particles anisotropy may give rise to non-trivial effects.
\textcite{Villarruel:2000} observed onset of nematic order in packing of
long rods in a narrow vertical tube subjected to vertical tapping.  The
rods initially compactify into a disordered state with predominantly
horizontal orientation, but at later times (after thousands of taps)
they align vertically, first along the walls, and then throughout the
volume of the pipe. The nematic ordering can be understood in terms of
the excluded volume argument put forward by \textcite{Onsager:1949}.

\textcite{Blair:2003} studied the dynamics of vibrated rods in a
shallow large aspect ratio system. Surprisingly, they found that
vertical alignment of rods at large enough filling fraction $n_f$
and the amplitude of vertical acceleration ($\Gamma>2.2$) can
occur in the bulk, and it does not require side walls. Eventually,
most of the rods align themselves vertically in a monolayer
synchronously jumping on the plate, and engage in a correlated
horizontal motion in the form of propagating domains of tilted rods,
multiple rotating vortices etc, see Fig. \ref{fig_blair1} and Fig.
\ref{fig_blair2}. The vortices exhibit almost rigid body rotation
near the core, and then the azimuthal velocity falls off, Fig.
\ref{fig_blair3}. The vortices merge in the course of their
motion, and eventually a single vortex is formed in the cell.

\begin{figure}
\caption{Phase diagram for the system of vertically-vibrated rods,
driving frequency 50 Hz. Vortices are observed for sufficiently
high filling fraction $n_f$ and above critical acceleration
$\Gamma$,  from \textcite{Blair:2003}} \label{fig_blair2}
\end{figure}

\begin{figure}
\caption{Azimuthal velocity of the vortex vs distance from the
center for different parameter values, from \textcite{Blair:2003}}
\label{fig_blair3}
\end{figure}

Experiments showed that the rod motion occurs when the rods are
tilted from the vertical, and it always occurs in the direction of
tilt. In subsequent work \textcite{Volfson:2004} experimentally
demonstrated that the correlated transport of bouncing rods is
also found in quasi-one-dimensional geometry, and explained this
effect using molecular dynamics simulations and a detailed
description of inelastic frictional contacts between the rods and
the vibrated plate. Effectively, bouncing rods become
self-propelled objects similar to other self-propelled systems,
for which large-scale coherent motion is often observed (bird
flocks, fish schools, chemotactic microorganism aggregation, etc.,
see e.g.
\textcite{Gregoire:2004,Helbing:2000,Helbing:2001,Toner:1995}).

\textcite{Aranson:2003} developed a phenomenological continuum
theory describing coarsening and vortex formation in the ensemble
of interacting rods. Assuming that
the motion of rods is overdamped due to the bottom friction,
the local horizontal
velocity ${\bf v}=(v_x,v_y)$  of rods is of  the form
\begin{equation}
{\bf v}= - \left(   \nabla p - \alpha   {\bf n} f_0(n) \nu
\right)/ \zeta \nu, \label{vel}
\end{equation}
where  $\nu$ is the density, $p$ is the hydrodynamic pressure,
the tilt vector ${\bf n}=(n_x,n_y)$ is the projection of the rod
director on the $(x,y)$ plane normalized by the rod length, i.e
$n=|{\bf n}|$,  and $\zeta$ is friction
coefficient.  According to \textcite{Blair:2003,Volfson:2004}, the
rods drift is determined by the average tilt of neighboring rods,
thus the term $\alpha {\bf n} f_0(n)\nu$ accounts for  the
average driving force from the vibrating bottom  on the tilted
rod. Eq. (\ref{vel}) combined with the mass conservation law
yields
\begin{equation}
\partial_t \nu = -{\rm div}  ( {\bf v} \nu)=\zeta^{-1}{\rm div}  \left( \nabla p - \alpha {\bf
n}f_0(n)  \nu\right). \label{mce1}
\end{equation}
To account for the experimentally observed phase separation and
coarsening \textcite{Aranson:2003} employed the Cahn-Hilliard
approach (see \cite{Bray:1994} for review) by assuming that
pressure $p$ can be obtained from the variation of a generic
bistable ``free energy'' functional $F$ with respect to the density
field $\nu$, $p = \delta F/ \delta \nu$.

To close the description the equation for the evolution of
tilt ${\bf n}$ is added on generic symmetry arguments:
\begin{eqnarray}
\partial_t {\bf n}&=& f_1 (\nu) {\bf n} -
|{\bf n}|^2 {\bf n} + \nonumber \\
&+&f_2 (\nu) \left ( \xi_1 \nabla^2 {\bf n}  + \xi_2 \nabla {\rm
div} {\bf n}  \right) + \beta \nabla \nu. \label{n1}
\end{eqnarray}
Here $f_{1,2}$ are certain functions of $\nu$,  $\xi_{1,2}$
characterize diffusion coupling between the neighboring rods.
Since the tilt field is not divergence-free, from the general
symmetry considerations both $\xi_{1,2} \ne 0$ \footnote{These
constants are analogous to the first and second viscosity in
ordinary fluids, see e.g. \cite{Landau:1959}}.

\begin{figure}
\caption{Sequence of images illustrating coalescence of vortices
in the model of vibrated rods, the field $|{\bf n}|$ is shown,
black dots corresponds to vortex cores where $|{\bf n}|=0$, from
\textcite{Aranson:2003}} \label{fig_rods1}
\end{figure}

Numerical and analytic studies of Eqs. (\ref{mce1}),(\ref{n1})
revealed phase coexistence, nucleation and coalescence of vortices
in accord with the experiment, see Fig. \ref{fig_rods1}.

An interesting experiment with anisotropic chiral particles was
performed by \textcite{Tsai:2005}.  The role of particles was played
by bend-wire objects which rotated in a preferred direction under
vertical vibration. The experiments demonstrated that individual
angular rotation of the particles was converted into the collective
angular momentum  of the granular gas of these chiral objects. The
theoretical description of this system was formulated in the
framework of two phenomenological equations for the density $\nu$
and center-of-mass momentum density $\nu {\bf v}$ and the spin
angular momentum density $l = {\cal I} \Omega$ arising from the
ration of particles around their center of mass, $\Omega$ is the
particle's  rotation frequency. Whereas the equations for density
and velocity are somewhat similar to those for the vibrated rod
system, the equation for the spin momentum clearly has no
counterpart in the vibrated rod system and was postulated in the
following form:
\begin{equation}
\partial_t l + \nabla {\bf v}l = \tau-\Gamma^\Omega-\Gamma (\Omega
- \omega) + D_\Omega \nabla^2 \Omega \label{omega1}
\end{equation}
where $\tau$ is the source of the angular rotation (due to
chirality of particles), $\omega$ is coarse-grained or collective
angular velocity, $\Gamma^\Omega $ and $\Gamma$ are dissipative
coefficients due to friction and $D_\Omega$ is the angular
momentum diffusion. Eq. (\ref{omega1}) predicts, in agreement with
the experiment, the onset of collective rotation of the gas of particles.
Possibly, it also exhibits non-trivial spatio-temporal dynamics
similar to those in the system of vibrated rods. However, due to
the small number of particles (about 350) in the experiment the
nontrivial collective regimes were not reported.

\subsection{Electrostatically driven granular media}
\label{subsec:electro}

Large ensembles of small particles display fascinating collective
behavior when they acquire an electric charge and respond to
competing long-range electromagnetic and short-range contact
forces. Many industrial technologies face the challenge of
assembling and separating such single- or multi-component micro
and nano-size ensembles. Traditional methods, such as mechanical
vibration and shear, are infective for very fine powders due to
agglomeration, charging, etc. Electrostatic effects often change
statistical properties of granular matter such as energy
dissipation rate \cite{Sheffler:2002}, velocity distributions
in granular gases \cite{Aranson:2002a,Kohlstedt:2005},
agglomeration rates  in suspensions \cite{Dammer:2004}, etc.

\textcite{Aranson:2000ed,Aranson:2002,Sapozhnikov:2003a,Sapozhnikov:2004}
studied electrostatically driven granular matter. This method
relies on the collective interactions between particles due to
a competition between short range collisions and long-range
electromagnetic forces. Direct electrostatic excitation of small
particles offers unique new opportunities compared to traditional
techniques of mechanical excitation. It enables one to deal with
extremely fine nonmagnetic and magnetic powders which are not
easily controlled by other means.

In most experimental realizations, several grams of mono-dispersed
conducting micro-particles  were placed into a 1.5 mm gap between
two horizontal $30\times30$ cm$^2$ glass plates covered by
transparent conducting layers of indium tin-dioxide. Typically
$45$ $\mu m$ Copper or $120$ $\mu m$ Bronze spheres were used.
Experiments were also performed with much smaller 1 $\mu m$
particles, \cite{Sapozhnikov:2004}.  An electric field
perpendicular to the plates was created by a  high voltage source
(0-3 kV) connected to the inner surface of each plate. Experiments
were performed in air, vacuum, or in the cell filled with
non-polar weakly-conducting liquid.

The basic principle of the electro-cell operation is as follows. A
particle acquires an electric charge when it is in contact with
the bottom conducting plate. It then experiences a force from the
electric field between the plates.  If the upward force induced by
the electric field exceeds gravity, the particle travels to the
upper plate, reverses charge upon contact, and is repelled down to
the bottom plate. This process repeats in a cyclical fashion. In
an air-filled or evacuated cell, the particle remains immobile at
the bottom plate if the electric field $E$ is smaller than the
first critical field $E_1$. For $E>E_1$ an isolated particle
leaves the plate and starts to bounce. However, if several
particles are in contact on the plate, screening of the electric
field reduces the force on individual particles, and they remain
immobile. A simple calculation shows that for the same value of
the applied electric field the force acting on isolated particles
exceeds by a factor of two the force acting on the particle inside
the dense monolayer. However, if the field is larger than a second
critical field value, $E_2> E_1$, all particles leave the plate,
and the system of particles transforms into an uniform gas-like
phase. When the field is decreased below $E_2 $\ ($E_1<E<E_2$), in
air-filled or evacuated cells  localized clusters of immobile
particles spontaneously nucleate to form a static clusters
(precipitate) on the bottom plate \cite{Aranson:2000ed}. The
clusters exhibit the Ostwald-type ripening
\cite{Meerson:1996,Bray:1994}, see also Subsec. \ref{sec:coars1}.

\subsubsection{Coarsening of clusters}
\label{sec:coars2}
 Results for the electrostatically driven system
yielded the following asymptotic scaling law, see Fig.
\ref{fig_sap2}:
\begin{equation}
N \sim \frac{1}{t} \label{law1}
\end{equation}
where $N$ is the number of clusters and $t$ is time. Accordingly,
the average cluster area $\langle A \rangle$ increases with time
as $\langle A \rangle \sim t$.  This behavior is consistent with
the {\it interface-controlled} Ostwald ripening \cite{Meerson:1996}.

A theoretical description of coarsening in an electrostatically driven
granular system was developed by \textcite{Aranson:2000ed},
\textcite{Sapozhnikov:2003}. The theory was formulated  in terms
of the Ginzburg-Landau-type equation for the number density of
immobile particles (precipitate or solid) $n$
\begin{equation}
\partial_t n= \nabla^2 n + \phi(n,n_g)
\label{gle1}
\end{equation}
where $n_g$ is the number density of bouncing particles  (gas)
$n_g$, and $\phi(n,n_g)$ is a function characterizing a solid/gas
conversion rate. The effectiveness of the solid/gas transitions is
controlled by the local gas concentration $n_g$. It was assumed
that the gas concentration is almost constant because the particle's
mean free pass in the gas state is very large. The gas
concentration $n_g$ is coupled to $n$ due  to total  mass
conservation constraint
\begin{equation}
S n_g+   \int \int n(x,y) dx dy= M, \label{mass_con}
\end{equation}
where $S$ is the area of domain of integration, and $M$ is the
total number of particles. Function $\phi(n,n_g)$ is chosen in
such a way as to provide bistable local dynamics of concentration
corresponding to the hysteresis of the gas/solid transition. The
above description yields a very similar temporal evolution of
clusters (see  Fig. \ref{fig_coars}) and produces a correct
scaling for the number of clusters Eq. (\ref{law1}).

\begin{figure}[ptb]
\caption{Illustration of phase separation and coarsening dynamics.
(a)-(c) Numerical solution of Eqs. (\ref{gle1}), (\ref{mass_con}),
white corresponds to dense clusters, black to dilute gas. (d)-(f)
show experimental results, from \textcite{Aranson:2002}.}
\label{fig_coars}
\end{figure}

In the so-called sharp interface limit when the size of clusters
is much larger than the width of interfaces between clusters and
granular gas, Eq. (\ref{gle1}) can be reduced to equations for the
cluster radii $R_i$ (assuming that clusters have circular form):
\begin{equation}
\frac{d R_i}{d t } = \kappa\left(\frac{1}{R_c(t)}  - \frac{1}{
R_i}\right), \label{radii}
\end{equation}
where $R_c$ is critical cluster size,  $\kappa$ is effective
surface tension (experimental measurements of cluster surface
tension were conducted by \textcite{Sapozhnikov:2003}). The
critical radius $R_c$ is a certain function of the granular gas
concentration $n_g$ that enters Eqs. (\ref{radii}) through the
the conservation law Eq. (\ref{mass_con}) which in two dimensions
reads
\begin{equation}
n_g S + \pi \sum_{i=1}^N R_i^2 = M. \label{con_mass2}
\end{equation}

The statistical properties of Ostwald ripening can be understood
in terms of the probability distribution function $f(R,t)$ of
cluster sizes. Following
\textcite{Lifsitz:1958,Lifsitz:1961,Wagner:1961}
and neglecting cluster merger, one obtains in the limit $N \to \infty$
that the probability
distribution $f(R,t)$ satisfies the continuity equation
\begin{equation}
\partial_t f + \partial_R \left( \dot R f \right) = 0.
\label{contin}
\end{equation}
From the mass conservation in the limit of small gas concentration Eq.
(\ref{con_mass2}) one obtains an additional constraint:
\begin{equation}
\pi \int_0^\infty R^2 f(R,t) d R = M \label{con_mass3}
\end{equation}
Eqs. (\ref{contin}),(\ref{con_mass3}) have a self-similar solution
in the form
\begin{equation}
f(R,t) = \frac{1}{t^{3/2} } F \left(\frac{R}{\sqrt t} \right)
\label{func1}
\end{equation}
For the total number of clusters $N=\int_0^\infty f d R$ the
scaling Eq.(\ref{func1}) yields $N \sim 1/t$, which appears to be
in a good agreement with the experiment, see Fig. \ref{fig_sap2}.
However, the cluster size distribution function appears to be in a
strong disagreement, see Fig. \ref{fig_sap3}. In particular,
\textcite{Lifsitz:1958,Lifsitz:1961,Wagner:1961} theory predicts
the distribution with a cut-off (dotted line) whereas the
experiment yields the function with an exponential tail.  A much
better agreement with the experiment was obtained when binary
coalescence of clusters was incorporated  in the
Lifshitz-Slyozov-Wagner theory
\cite{Sapozhnikov:2005,Conti:2002}.The coalescence events become
important for a finite area fraction of the clusters.

\begin{figure}
\caption{Average cluster area $\langle A(t) \rangle $ (a) and
inverse number of clusters $1/N(t)$ vs time in air-filled cell.
The straight line in (b) shows theoretical prediction $1/N \sim
t$, from \textcite{Sapozhnikov:2005}} \label{fig_sap2}
\end{figure}

\begin{figure}
\caption{Scaled cluster size distribution function $F(\xi)$ with
$\xi=R/\sqrt t$. The squares show experimental results, the dotted
line shows analytic result form Lifshitz-Slyozov-Wagner theory
\cite{Wagner:1961}, and solid line shows $F$ obtained from the
theory accounting for binary coalescence,  from
\textcite{Sapozhnikov:2005}} \label{fig_sap3}
\end{figure}

\textcite{Bennaim:2003} applied an exchange growth model to
describe coarsening in granular media. In this theory the cluster
growth rates are controlled only by the cluster area ignoring
shape effects. Assuming that the number of particles  in a cluster
evolves via uncorrelated exchange  of single particles with an other
cluster the following equation for the density of clusters
containing $k$ particles can be derived:
\begin{equation}
\frac{d A_k}{dt}= \sum_{i,j} A_i A_j K_{ij} \left( \delta_{k,i+1}+
\delta_{k,i-1}-2 \delta_{k,i}\right) \label{bennaim1}
\end{equation}
where $A_k$ is the probability to find a cluster containing $k$
particles, $K_{ij}$ the exchange kernel and $\delta_{k,i}$ is the
Kronecker symbol. For the choice of homogeneous kernel $K_{ij}=(i
j)^\lambda$ with $\lambda=1$ this theory predicts correct scaling
of the cluster size with time $R \sim \sqrt t$ and exponential
decay of the cluster size distribution function, as in the
experiment. The choice of $\lambda=1$ is equivalent to the
assumption that the exchange rate is determined by the size of the
cluster. In the theory by \textcite{Sapozhnikov:2005}  the cluster
evolution is governed by the evaporation/deposition of particles
at the interface of the cluster and controlled by the overall
pressure of the granular gas. Thus, both theories predict the same
scaling behavior, however the underlying assumptions are very
different. A possible explanation for this may be that while the
exchange growth model ignores the curvature of the cluster
interface and the dependence on exchange rate on the pressure of
granular gas, the agreement is obtained by tuning the adjustable
parameter $\lambda$.

\subsubsection{Dynamics of patterns in a fluid-filled cell}

\textcite{Sapozhnikov:2003a} performed experiments with
electrostatically driven granular media immersed in a weakly
conducting non-polar fluid (toluene-ethanol mixture). Depending on
the applied electric field and the ethanol concentration (which
controls the conductivity of the fluid), a plethora of static and
dynamic patterns were discovered, see Fig. \ref{fig_sap4}. For
relatively low concentrations of ethanol  (below 3\%), the
qualitative behavior of the liquid-filled cell is not very
different from that of the air-filled cell: clustering of
immobile particles and coarsening were observed between two
critical field values $E_{1,2}$ with the clusters being
qualitatively similar to that of the air cell. However, when the ethanol
concentration is increased, the phase diagram becomes asymmetric
with respect to the direction of the electric field.  Critical
field magnitudes, $E_{1,2}$, are larger when  the electric field is
directed downward (``$+$" on the upper plate) and smaller when the field
is directed upward (``$-$" on the upper plate). This difference
increases with ethanol concentration. The observed asymmetry of
the critical fields is apparently due to an excess negative charge
in the bulk of the liquid.

The situation changes dramatically for higher ethanol
concentrations: increasing the applied voltage leads to the
formation of two new immobile phases: honeycomb (Fig.
\ref{fig_sap4}b) for the downward direction of the
applied electric field, and two-dimensional crystal-type states for the
upward direction.

A further increase of ethanol concentration leads to the
appearance of a novel dynamic phase - condensate (Fig.
\ref{fig_sap4}c,d) where almost all particles are engaged in a
circular vortex motion in the vertical plane, resembling
Rayleigh-B\'enard convection. The condensate co-exists with the
dilute granular gas. The direction of rotation is determined by
the polarity of the applied voltage: particles stream towards the
center of the condensate near the top plate for the upward field
direction and vice versa. The evolution of the condensate
depends on the electric field direction. For the downward field, large
structures become unstable due to the spontaneous formation of
voids  (Fig. \ref{fig_sap4}d). These voids exhibit complex
intermittent dynamics.  In contrast, for the upward field, large
vortices merge into one, forming an asymmetric object which often
performs composite rotation in the horizontal plane. The pattern
formation in this system is most likely caused by self-induced
electro-hydrodynamic micro-vortices created by the particles in
weakly-conducting fluids. These micro-vortices create long-range
hydrodynamic vortex flows which often overwhelm electrostatic
repulsion between likely-charged particles and introduce
attractive dipole-like hydrodynamic interactions. Somewhat similar
micro-vortices are known in driven colloidal systems, see e.g.
\cite{Yeh:1997}.

\textcite{Aranson:2004} developed a phenomenological continuum
theory of pattern formation for metallic micro-particles in a
weakly conducting liquid  subject to  an electric field. Based on the
analogy with  the previously developed theory of coarsening in
air-field cell \cite{Aranson:2002}, the model is formulated in
terms of conservation laws for the number densities of immobile
particles (precipitate) $n_p$ and bouncing particles (gas) $n_g$
averaged over the thickness of the cell:
\begin{eqnarray}
\partial_t n_p = \nabla {\bf J}_p + f \;,\;
\partial_t n_g = \nabla {\bf J}_g - f.
\label{con_laws}
\end{eqnarray}
Here $J_{p,g}$ are the mass fluxes of precipitate and gas
respectively and the function $f$ describes gas/precipitate
conversion which depends on $n_{p,g}$, electric field $E$ and
local ionic concentration $c$. The fluxes are written as:
\begin{equation}
{\bf J}_{p,g}=D_{p,g} \nabla n_{p,g} + \alpha_{p,g}(E)  {\bf
v}_\perp  n_{p,g} (1- \beta(E)  n_{p,g}), \label{flux}
\end{equation}
where $v_\perp$ is horizontal hydrodynamic velocity, $D_{p,g}$ are
precipitate/gas diffusivities. The last term, describing particles
advection  by fluid, is reminiscent of the Richardson-Zaki relation
for a drag force frequently used in the engineering literature
\cite{Richardson:1954}. The factor $(1- \beta(E) n_{p,g})$
describes the saturation of flux at large particle densities $n
\sim 1/\beta$ due to the decrease of void fraction. Terms $\sim
\alpha_{p,g}$ describe advection of particles by the fluid.
Interestingly, in the limit of very large gas diffusion $D_g \gg
D_p$ and without advection terms ($\alpha_{p,g}=0$) the model
reduces to Eqs. (\ref{gle1}) and (\ref{mass_con}) applied for
air-filled cell \cite{Aranson:2002}.

Eqs. (\ref{con_laws}) are coupled to the cross section averaged
Navier-Stokes equation for vertical velocity $v_z$:
\begin{equation} n_0 (\partial_t v_z +
{\bf v} \nabla v_z)= \mu \nabla^2 v_z -\partial_z p + E_z q
\label{nse}
\end{equation}
where $n_0$ is the density of liquid (we set $n_0=1$), $\mu$ is
the viscosity, $p$ is the pressure, and $q$ is the charge density.
The last term describes the electric force acting on charged
liquid. Horizontal velocity $v_\perp$  is obtained from $v_z$
using the incompressibility condition $\partial_z v_z +
\nabla_\perp v_\perp = 0 $ in the approximation that vertical
vorticity $\Omega_z=\partial_x v_y-\partial_y v_x$  is small
compared to in-plane vorticity. This assumption allows one to find
the horizontal velocity as a gradient of quasi-potential $\phi$:
${\bf v}_\perp=- \nabla_\perp \phi$.

\begin{figure}
\caption{Sequence of snapshots illustrating evolution of pulsating
rings  (top raw) and rotating vortices (bottom raw) obtained from
numerical solution of Eqs. (\ref{con_laws}),(\ref{nse}), from
\textcite{Aranson:2004}} \label{fig_ar1}
\end{figure}

For an appropriate choice of the parameters the model Eqs.
(\ref{con_laws}),(\ref{nse}) yields qualitatively correct phase
diagram and the patterns observed in the experiment, see Figs.
\ref{fig_sap4} and \ref{fig_ar1}.

\subsection{Magnetic particles}

Electric and magnetic interactions allow introduction of
controlled  long-range forces in granular systems.
\textcite{Blair:2003,Blair:2003b,Stambaugh:2004a,Stambaugh:2004b}
performed experimental studies with vibrofluidized magnetic
particles. Several interesting phase transitions were reported, in
particular, the formation of dense two-dimensional clusters and
loose quasi-one-dimensional chains and rings.
\textcite{Blair:2003} considered pattern formation in a mixture of
magnetic and non-magnetic (glass)  particles of equal mass. The
glass particles played the role of ``phonons", their concentration
allowed an adjustment of the typical fluctuation velocity of the
magnetic subsystem. The phase diagram delineating various regimes
in this system is shown in Fig. \ref{fig_blair4}. While the phase
diagram shows some similarity with equilibrium dipolar fluids
(such as phase coexistence), most likely there are differences due
to the non-equilibrium character of granular systems.

\begin{figure}
\caption{Phase diagram illustrating various regimes in magnetic
granular media, $T$ is the temperature determined from the the
width of  velocity distribution, $\Phi$ is surface coverage
fraction of glass particles,  from \textcite{Blair:2003b}}
\label{fig_blair4}
\end{figure}

\textcite{Stambaugh:2004a} performed experiments with relatively
large particles (about 1.7 cm), and near the  closed-packed
density. It was found that particles form hexagonal closed-packed
clusters in which the magnetic dipoles lay in the plane and assume
circulating vortical patterns. For lower density ring patterns
were observed. Experiments with mixture of particles with two
different magnetic moments revealed segregation effects
\cite{Stambaugh:2004b}. The authors argue that the static
configurational magnetic energy is the primary factor in pattern
selection.

Experiments by
\textcite{Blair:2003b,Stambaugh:2004a,Stambaugh:2004b} were
limited to a small number (about 10$^{3}$) of large  particles due
to the intrinsic limitation of the mechanical vibrofluidization
technique. \textcite{Snezhko:2005} performed experimental studies
of 90 $\mu m$  Nickel micro-particles subjected to electrostatic
excitation, see also Subsec. \ref{subsec:electro}. The
electrostatic system allowed researchers to perform experiments
with a very large number of particles (of the order of $10^6$) and
a  large aspect ratio of the experimental cell. Thus the
transition between small chains and large networks (Fig.
\ref{fig_snezhko}) was addressed in detail. An abrupt divergence
of the chain length was found when the frequency of field
oscillations decreased, resulting in the formation of a giant
interconnected network.

Studies of the collective dynamics and pattern formation of
magnetic particles are still in the early phases.  While it is
natural to assume that magnetic interaction plays a dominant role
in pattern selection, further computational and theoretical
studies of pattern formation in systems of driven dipolar
particles are necessary. Besides a direct relevance for the
physics of granular media, studies of magnetic granular media may
provide an additional insight into the behavior of dipolar hard
sphere fluids where the nature of solid/liquid transitions is
still debated \cite{deGennes:1970,Levin:1999}. Vibration or
electrostatically fluidized magnetic particles can also be viewed
as a macroscopic model of a ferrofluid, where similar experiments
are technically difficult to perform.

\section{Overview and Perspectives}
\label{sec:over}

Studies of granular materials are intrinsically interdisciplinary
and they borrow ideas and methods from other fields of physics
such as statistical physics, mechanics, fluid dynamics, and the theory of
plasticity. On the flip side, progress in understanding granular
matter can be often applied to seemingly unrelated physical
systems, such as ultra-thin liquid films, foams, colloids,
emulsions, suspensions, and other soft condensed matter systems.
The common feature shared by these systems is the discrete
microstructure directly influencing macroscopic behavior.
For example, the order
parameter description similar to that of Sec.
\ref{subsubsec:part} was applied to stick-slip
friction in ultra-thin films, \cite{Israelachvili:1988,Urbach:2004,Carlson:1996,Aranson:2002c}.

\textcite{Lemaitre:2002,Lemaitre:2004} applied the idea of
shear-transformation zone (STZ) pioneered by \textcite{Falk:1998}
for amorphous solids both to granular matter and to the boundary
lubrication problem in confined fluid. In this theory the plastic
deformation is represented by a population of mesoscopic regions
which may undergo non-affine deformations in response to stress.
Concentration of STZs in amorphous material is somewhat similar
to the order parameter (relative concentration of defects)
introduced by \textcite{Aranson:2002c}. A conceptually similar approach
was proposed by \textcite{Staron:2002} who described the onset
of fluidization as a percolation of the contact
network with fully mobilized friction.  Whereas derivation of the
constitutive relations from first-principle microscopic rules is
still a formidable challenge, these approaches are promising for
understanding of not only the boundary lubrication problem, but also
onset of motion in dense granular matter.

Flowing liquid foams and emulsions  share many similarities with
granular matter: they have internal discrete structure (bubbles
and drops  play the role of grains), and two different mechanisms are
responsible for the transmission of stresses:  elastic for small stress
and visco-plastic above certain yield stress. However, there are
additional complications: bubbles are highly  deformable and,
unlike granular matter, a number of particles may change due to
the coalescence of bubbles.

Foams and granular materials often exhibit similar behavior, such
as non-trivial  stress relaxation and power-law distribution of
rearrangement events \cite{Dennin:1997}.  Stick-slip behavior was
reported both for sheared foams \cite{Lauridsen:2002} and granular
materials \cite{Nasuno:1997}. Remarkably, recent experiments with
two dimensional foams \cite{Lauridsen:2004} and three dimensional
emulsions  \cite{Coussot:2002a,Coussot:2002b,DaCruz:2002} strongly
suggest the coexistence between flowing (liquid) and jammed
(solid) states reminiscent of that in granular matter. Furthermore,
avalanche behavior reminiscent of granular flows down an inclined
plane \cite{Daerr:1999} was reported by \textcite{Coussot:2002a}
for clay suspensions, see Fig. \ref{fig_coussot}. There are many
approaches treating foams, gel and suspensions as complex fluids
with specific stress-strain constitutive relation. For example,
\textcite{Fuchs:2002} used the analogy between glasses and dense
colloidal suspensions and applied the mode coupling approach to
understand the nonlinear rheology and yielding. Similar approaches
can be possibly useful for granular materials \cite{Schofield:1994}.

\begin{figure}
\caption{Sequence of snapshots illustrating evolution of clay
suspension drop poured over sandpaper, from
\textcite{Coussot:2002a}} \label{fig_coussot}
\end{figure}


\textcite{Liu:1998} suggested that a broad class of athermal soft
matter systems (glasses, suspensions, granular materials) shows a
universal critical behavior in the vicinity of solid-fluid or {\em
jamming} transition, see Fig. \ref{fig_liu}. Whether jammed
systems indeed have common features that can be described by a
universal phase diagram is an open issue. An interesting question
in this context is a possibility of thermodynamic description of
driven, macroscopic, athermal systems like granular materials and
foams in terms of some kind of effective temperature. Studies of
interacting particles under shear
\cite{Ono:2002,Ohern:2004,Makse:2002,Xu:2005,Corvin:2005} indicate
that indeed under certain conditions it is possible to define an
effective temperature (for example, from the equivalent of the
Einstein-Stokes relation) for a broad class of athermal systems
from comparison of the mechanical linear response with the
corresponding time-dependent fluctuation-dissipation relation.
However, the possibility of developing nonequilibrium
thermodynamics of the basis  of the effective temperature is
under debate.

\begin{figure}
\caption{A possible phase diagram for jamming. The jammed region,
near the origin, is enclosed by the depicted surface. The line in
the temperature-load plane is speculative, and indicates how the
yield stress might vary for jammed systems in which there is
thermal motion, from \textcite{Liu:1998}} \label{fig_liu}
\end{figure}

Granular systems exhibit many similarities with traffic flows and
collective motion of self-propelled particles such as swimming bacteria, fish
schools, bird flocks, etc., see for review \cite{Helbing:2001}. In
particular, jamming transition in granular media and traffic jams
show similar features, such as hysteresis, and clusters formation.
Moreover, continuum models of traffic flows are often cast in the
form of modified Navier-Stokes equation with density-dependent
viscosity, similar to granular hydrodynamics.

Let us discuss briefly some open questions in the physics of
granular matter.

\begin{itemize} \item Static vs. dynamic description. Commonly accepted
models of rapid granular flows (granular hydrodynamics) and
quasi-static dense flows (elastic and visco-plastic models) are
very different, see e.g. \textcite{Goldenberg:2002}. However, near
the fluidization transition, and in dense partially-fluidized
flows, the differences between these two regimes become less
obvious. The fluidization of sheared granular materials has many
features of a first-order phase transition. The phenomenological
partial fluidization theory in principle can be a bridge between
the static and dynamic descriptions. The order parameter related
to the local coordination number appears to be one of the hidden
fields required for a consistent description of granular flows.
One important question in this regard is the universality of the
fluidization transition in different granular systems and
geometries. On the opposite side of the fluidization transition,
the static state of the granular matter can be described by the
order parameter related to the percentage of static contacts with
fully activated dry friction (critical contacts)
\cite{Staron:2002}. It was shown that once these contacts form a
percolation cluster, the granular pack slips and fluidization
occurs. It is of obvious interest to relate this ``static'' order
parameter and the ``dynamics'' order parameter discussed above. We
see one of the main future challenges in the systematic derivation
of the continuum theory valid both for flowing and static granular
matter.

\item
Statistical mechanics of dense granular systems.
Clearly, discrete grain structure plays a major role in the dynamics and
inherent stochasticity of granular response. The number of particles in a
typical granular assembly is large (10$^6$ or more) but it is much
smaller than the Avogadro number. Traditional tools of statistical physics do
not apply to dense granular systems since grains do not exhibit thermal Brownian
motion.  One of the alternative ways of describing statistics of
granular media was suggested by \textcite{Edwards:1998} in which they
proposed that volume rather than energy serves as the extensive
variable in a static granular system, so that the role of temperature is
played by the compactivity which is the derivative of the volume with
respect to the usual entropy. Recent experiments
\cite{Makse:2002,Schroter:2005} aim to test this theory experimentally.
Connecting Edwards theory with granular hydrodynamics will be an
interesting challenge for future studies.


\item Realistic simulations of three-dimensional granular flows.
Even the most advanced simulations of granular flows in three
dimensions \cite{Silbert:2003,Silbert:2005} are limited to
relatively small samples (e.g. $100\times40\times40$ particles
box) and are very time consuming. The granular problems are
inherently very stiff: while the collisions between particles are
very short ($O(10^{-4}sec$), the collective processes of interest
may take many seconds or minutes. As a result, to the time step
limitations a simulation of realistic hard particles is not
feasible: the ``simulations'' particles have elastic moduli
several orders of magnitude smaller than sand or glass. The
particle softness may introduce unphysical artifacts in the
overall picture of the motion. Different approaches to handling
this problem will be necessary to advance the state of the art in
simulations. New opportunity can be offered by the equation-free
simulation method proposed by \textcite{Kevrekidis:2004}. Another
area of simulations which needs further refinements is an accurate
account of dry friction. In the absence of a better solution
current methods (see for review \cite{Luding:2004}) employ various
approximate techniques to simulation dry friction, and accuracy of
these methods can be questionable.

\item
Complex interactions. Understanding of dynamics of granular systems with complex interactions
is certainly an intriguing and rapidly developing field.  While
interaction of grains with intersticial fluid is a traditional part of
engineering research, effects of particle anisotropy, long-range
electromagnetic interactions mediating collisions, adhesion,
agglomeration and many others constitute a formidable challenge for
theorists and a fertile field of future research.

\item Granular physics on a nano-scale. There is a persistent
trend in the industry such as powder metallurgy, pharmaceutical
and various chemical technologies towards operating with smaller
and smaller particles.  Moreover, it was recognized recently that
micro- and nano-particles can be useful for fabrication of desired
ordered structures and templates for a broad range of
nanotechnological applications through self-assembly processes.
Self-assembly, the spontaneous organization of materials into
complex architectures, constitutes one of the greatest hopes of
realizing the challenge to create ever smaller nanostructures. It
is a particulary attractive alternative to traditional approaches
such as lithography and electron beam writing. Reduction of the
particle size to micro- and nano -scales shifts the balance
between forces controlling particle interaction because the
dominant interactions depend on the particle size. While for
macroscopic grains the dynamics are governed mostly by the
gravity, collisional and frictional forces, for micro- and
nano-particles the dominant interactions include long-range
electromagnetic forces, short- range van der Waals interactions,
etc. Nevertheless, some concepts and ideas developed in the
``traditional'' granular physics  were successfully applied to
understand dynamic self-assembly of microparticles
\cite{Sapozhnikov:2003a,Sapozhnikov:2004} and even biological
microtubules \cite{Aranson:2005}. We expect to see more and more
efforts in this direction.

\end{itemize}

\section*{Acknowledgments}

The authors  thank  Dmitrii Volfson, Alexey Snezhko, Maksim
Saposhnikov, Jie Li, Adrian Daerr,  Bob Behringer, Jerry Gollub,
Thomas Halsey, Denis Ertas, Harry Swinney, Jeff Olafsen, Eli
Ben-Naim, Valerii Vinokur, Wai Kwok, George Crabtree, Paul
Umbanhowar, Francisco Melo, Eric Clement, Jacques Prost, Philippe
Claudin, Julio Ottino, Devang Khakhar, Jean-Philippe Bouchaud,
Olivier Pouliquen, Jacques Duran, Ana\"el  Lema\^itre, Evelyne
Kolb, Hugues Chat\'e, Gary Grest, Arshad Kudrolli, Douglas Durian,
Peter Schiffer, Leo Silbert, Wolfgang Losert, Daniel Blair, Paul
Chaikin, Henrich Jaeger, Sid Nagel, Leo Kadanoff, Thomas Witten,
Sue Coppersmith, Baruch Meerson, Ray Goldstein, Chay Goldenberg,
Isaak Goldhirsch, Robert Ecke, Thorsten P\"oschel, Alexandre
Valance, James Dufty, James Jenkins,  Dietrich Wolf, Haye
Hinrichsen, Lorenz Kramer, Len Pismen, Martin van Hecke, Wim van
Saarloos, Guenter Ahlers, Jacob Israelachvili, James Langer,
Pierre-Gilles de Gennes and many others for useful discussions.
This work was supported by the Office of the Basic Energy Sciences
at the United States Department of Energy, grants W-31-109-ENG-38,
and DE-FG02-04ER46135.

The review was partly written when one of us (I.A) was attending
Granular Session in Institute Henry Poincar\'e, Paris,  and
Granular Physics Program, Kavli Institute for Theoretical Physics
in Santa Barbara.

\bibliographystyle{apsrmp}

\end{document}